\documentclass[12pt]{iopart}
\usepackage{cite}
\usepackage[english]{babel}
\expandafter\let\csname equation*\endcsname\relax
\expandafter\let\csname endequation*\endcsname\relax
\usepackage{amsmath}
\usepackage{amsfonts,amssymb}
\usepackage{color,soul}
\usepackage{hyperref}
\usepackage{listings}
\usepackage{graphicx,subfigure}

\begin{document}

\title[Plasma chemistry with LoKI-GM]{LoKI-GM: a global model framework for plasma chemistry studies}

\author{L.~L.~Alves$^1$, A.~Tejero-Del-Caz$^2$, L. Marques$^3$, P. Pereira$^1$, N. Pinhão$^1$, C.~D. Pintassilgo$^{4,1}$, T. Silva$^1$, P. Viegas$^1$ and V. Guerra$^1$}

\address{$^1$Instituto de Plasmas e Fusão Nuclear, Instituto Superior Técnico, Universidade de Lisboa, Lisbon, Portugal}
\address{$^2$Departamento de Física, Universidad de Córdoba, Cordoba, Spain}
\address{$^3$ Centro de Física das Universidades do Minho e do Porto, Universidade do Minho, Braga, Portugal}
\address{$^4$Faculdade de Engenharia, Universidade do Porto, Porto, Portugal}
\ead{llalves@tecnico.ulisboa.pt}
\vspace{10pt}
\begin{indented}
\item[] \today
\end{indented}

\begin{abstract}
Global (zero-dimensional or spatially averaged) models are widely employed to study complex chemistries in low-temperature plasmas (LTPs). By adopting a spatially average description of the plasma, they substantially reduce the computational cost while still providing reliable and detailed insight into the key processes taking place in the plasma. In their most general formulation, global models involve the coupled solution of a Boltzmann solver (to describe the electron kinetics) and a Chemistry solver (to describe the heavy-species kinetics).  
This paper presents a tutorial on the LisbOn Kinetics Global Model (LoKI-GM) framework, developed in MATLAB and available as open-source code. The framework couples the Boltzmann solver LoKI-B, which solves the space-independent form of the two-term electron Boltzmann equation for non-magnetised non-equilibrium LTPs, excited by DC/HF electric fields or time-dependent (non-oscillatory) electric fields, and the Chemistry solver LoKI-C, which solves the system of zero-dimensional rate balance equations for the main charged and neutral species in the plasma and at the surface, receiving as input the kinetic schemes for the gas/plasma/surface system under study. The inclusion of several transport models together with support for surface kinetics models are distinguishing features of LoKI-GM compared with other global chemistry models. 
We briefly present the formulation of LoKI-GM, including its numerical solution strategy, input/output parameters, and calculation workflow for both active discharges and afterglow plasmas. The main differences with respect to existing global models are highlighted, and the flexibility of the code for plasma chemistry studies is demonstrated through representative simulation results obtained across a range of gas discharge configurations and operating conditions.
\end{abstract}

\ \\
\ \\
%
\submitto{\PSST}

%
\maketitle
%
%

\section{Introduction}

Modelling activities are a fundamental component in any research domain, complementing and/or aiding experimental diagnostics; providing predictions on the behaviour of significant quantities, especially when experimental access is limited; and contributing to a deeper understanding of the field's fundamental knowledge. Modelling and simulation of low-temperature plasmas (LTPs) have been considered as requirements for advancing the field, and model-based design for plasma equipment and processes has been identified as a critical capability realizing industrial objectives~\cite{Adamovich2022}. The guidance offered by LTP models proves particularly valuable given the inherent complexity of the medium (often characterized by different material phases), composed by charged particles (electrons and ions) and by neutral species in different excited states, intrinsically in non-equilibrium as a result of collisional, radiative and electromagnetic interactions.

Global (zero-dimensional or spatially averaged) models are widely employed to study complex chemistries in LTPs. By adopting a spatially averaged description of the plasma, these models substantially reduce the computational cost while still enabling the identification of dominant reaction pathways and the creation and destruction mechanisms of the main plasma species. In their most general formulation, global models involve the coupled solution of an electron Boltzmann solver, to describe the electron kinetics, and a Chemistry solver, to account for the kinetics of the heavy species. However, the effective use of these models depends strongly on the details of their implementation, designed to mitigate the intrinsic limitations of the zero-dimensional approximation and to broaden their practical applicability. These limitations are typically associated with the relevant pressure regimes and their implications for a self-consistent treatment of the electron kinetics, as well as for the handling of space-charge sheaths and spatial density gradients.
A comprehensive review of global models is presented by Alves \textit{et al.}~\cite{Alves2018}, along with an extensive list of references for complementary in-depth reading. In the following, only the main aspects of these models are highlighted.

At sufficiently high pressures (e.g., above $\sim 100$~Pa)~\cite{Ingold1997}, electrons undergo frequent collisions and may be considered in equilibrium with the local reduced electric field $E/N$ (where $E$ is the electric-field amplitude and $N$ is the gas density), so that the local field approximation (LFA) is valid. Under these conditions, the electron kinetics can be described self-consistently by an electron energy distribution function (EEDF) obtained from a Boltzmann solver, for example using the homogeneous two-term approximation for $E/N \stackrel{<}{\sim} 250$~Td. In this pressure regime, charge separation effects are typically confined to thin sheath regions near the boundaries, and can often be neglected in spatially averaged formulations. 
In the absence of transport (including gas flow), the plasma dynamics is then predominantly governed by chemical reactions, which locally modify the plasma composition and properties. In this context, \textit{global zero-dimensional} models can be applied locally at different spatial positions, effectively becoming \textit{local non-dimensional} models. Within this generalization, 0D models have been used to describe strongly non-homogeneous systems by partially accounting for spatial effects through the analogy between batch reactors and plug-flow reactors. In this framework, temporal variations are mapped onto spatial variations as a function of the distance traveled by the gas at a given flow rate. This approach has been successfully applied, for example, to reproduce the experimental conditions of atmospheric-pressure plasma jets~\cite{Gaens2013}, to model reactors for gas conversion~\cite{Alves2018}, and to account for the filamentary behaviour of dielectric barrier discharges~\cite{Aerts2012,Snoeckx2013,Bogaerts2016}.

Conversely, at very low pressures (e.g., a few Pa), the electron energy relaxation length exceeds the characteristic plasma dimensions, and the total electron energy (kinetic plus potential) remains approximately constant during the electron motion. As a result, the EEDF becomes strongly non-local, and the electron Boltzmann equation (EBE) must be solved using a non-local formulation~\cite{Bernstein1954,Tsendin1974}, unless the ionization degree is sufficiently high (e.g., above $\sim 10^{-2}$), in which case electron–electron collisions dominate and a Maxwellian EEDF may be assumed. In this low-pressure regime, space-charge sheaths become highly extended and can occupy a significant fraction of the discharge volume, thereby strongly influencing the transport of charged species. Consequently, despite the relatively homogeneous spatial profiles of species densities and energy, global models are generally not the most appropriate modeling approach under these conditions. 
At intermediate pressures (typically $\sim 10$–$100$~Pa)~\cite{Ingold1997}, the local mean energy approximation (LEA) applies~\cite{Ingold1997,Alves2007,Meijer1992,Dias2025}. In this regime, the self-consistent coupling between Boltzmann and Chemistry solvers should be performed through the electron mean energy $\varepsilon$, obtained from the corresponding energy balance equation, rather than directly through the reduced electric field. Although space-charge sheaths and density gradients can still influence species transport at intermediate pressures, several spatially averaged formulations have been proposed in the literature to incorporate these effects within global models~\cite{Alves2023}.

Therefore, owing to their spatially averaged formulation, global (0D) models are best suited to intermediate- and high-pressure regimes, where the effects of space-charge sheaths and density gradients can either be neglected or adequately represented through averaged transport descriptions.

Global models can be solved using several codes available in the LTP community, for example: ZDPlasKin, a freeware code developed by Pancheshnyi \textit{et al.}~\cite{Pancheshnyi2008}; GlobalKin, developed by Kushner \textit{et al.}~\cite{Lietz2016,Stafford2004} and available upon request; Quantemol-P, developed with a graphical user interface~\cite{Munro2008}, corresponding to a commercial application that extends GlobalKin to the automatic generation of a plasma chemistry from a set of user-defined species; the global model within the PLASIMO commercial software, developed by van Dijk \textit{et al.}~\cite{Dijk2009}; CRANE (Chemical ReAction NEtwork), a MOOSE-based open-source tool for plasma chemistry applications, developed in order to add an independent chemical kinetics application to the framework~\cite{Keniley2019,CRANE}; the Kinetic Global Model framework (KGMf), a global model simulation code developed to explore the reaction kinetics and pathways in plasma discharge systems~\cite{Krek2021,KGMf} and available upon request; ChemPlasKin, an open-source code for zero-dimensional simulations that computes the time-resolved evolution of species concentration and gas temperature in a unified gas–plasma kinetics framework~\cite{Shao2024,ChemPlasKin}. 

This work presents a tutorial on the LisbOn Kinetics Global Model (LoKI-GM) for plasma chemistry studies. The framework, available as open-source software~\cite{LoKISuite,LoKISuiteGit}, solves the global kinetic model for pure gases and gas mixtures, providing a self-consistent description of the chemical kinetics and transport of charged and neutral species, in both the volume and surface phases, for user-defined operating conditions: gas composition mixture, pressure, reactor dimensions, and excitation modes (DC/HF glow discharges, post-discharge, and pulsed operation). LoKI-GM comprises two modules, that can run self-consistently coupled or as standalone tools:
\begin{description}
\item[(i)] a Boltzmann solver (LoKI-B)~\cite{Tejero2019,Tejero2021}, which solves the space independent form of the two-term EBE to calculate the isotropic and the anisotropic parts of the electron distribution function (the former corresponding to the EEDF), and the associated electron macroscopic parameters. LoKI-B applies to non-magnetised non-equilibrium LTPs, excited by DC/HF electric fields or time-dependent (non-oscillatory) electric fields from different gases or gas mixtures. The tool uses a stationary description for DC fields, a Fourier time-expansion description for HF fields, and a time-dependent description for time-varying fields. LoKI-B describes first and second-kind electron collisions with electronic, vibrational and rotational states, includes electron-electron collisions, and handles rotational collisions using either a discrete formulation or a continuous approximation. It also accounts for variations in the electron population caused by non-conservative events (ionization and attachment) by adopting temporal or spatial growth models for the electron density;
\item[(ii)] a Chemical solver (LoKI-C), which solves the system of zero-dimensional (volume average) rate balance equations for the main charged and neutral species in the plasma. LoKI-C receives as input data the kinetic schemes for the gas/plasma/surface system under study, via an intuitive csv-like input file, and gives as output the particle densities of the different gas/plasma/surface species, the corresponding creation/destruction reaction rates, and the reduced electric field (and any related quantity, such as the discharge current or the discharge power-density). The tool uses several modules to describe the mechanisms (collisional, radiative and transport) controlling the creation/destruction of species, includes a gas/plasma thermal model for the self-consistent calculation of the gas temperature, and supports multi-component mean-field microkinetic mesoscopic models to handle surface kinetics in a fully coupled way with volume kinetics. Note that the inclusion of several transport models and support for surface kinetics models (see sections~\ref{General-rateCoeffVolSurf} and~\ref{General-rateCoeffTransp}) are distinguishing features of LoKI-C compared with other global chemistry models.
\end{description}

LoKI-GM embodies the software implementation of the expertise in plasma-chemistry modelling accumulated over many years by the N-PRiME group in Lisbon~\cite{nprime}. Its development has been guided by four main principles: (i) \textit{unbundling} of components, based on an ontology that prioritizes a clear separation between tool and data; (ii) \textit{flexibility} to host any plasma chemistry scheme, covering both volume and surface mechanisms and allowing a full state-to-state description; (iii) \textit{consistency} in the coupling between electron kinetics and heavy-species kinetics, particularly in determining the discharge characteristic; and (iv) \textit{clarity} in defining the working conditions relevant for comparing simulations results with experimental measurements. LoKI-GM is implemented using a flexible and extensible object-oriented architecture in MATLAB~\cite{MATLAB}, leveraging its matrix-based framework and high-performance ODE solvers. The code is distributed under the GNU General Public License 3.0.

While preparing its release as open-source code, the framework has been successfully applied to various plasma systems, including DC glow discharges sustained in O$_2$~\cite{Dias2023,Viegas2024,Viegas2025}, CO$_2$~\cite{Ogloblina2021,Guerra2022,Liu2025,Viegas2026}, CO$_2$-N$_2$~\cite{Fromentin2023}, CO$_2$-O$_2$~\cite{Fromentin2023a} and CO$_2$-CH$_4$~\cite{Baratte2024}; inductively coupled discharges sustained in O$_2$~\cite{Annusova2018}; DC glow discharges, radio-frequency discharges and afterglows in N$_2$–O$_2$~\cite{Guerra2019,Silva2024}; microwave discharges sustained in N$_2$–O$_2$~\cite{Coche2016}; pulsed glow discharges sustained in O$_2$~\cite{Marinov2013}, N$_2$-O$_2$~\cite{Pintassilgo2019} and CO$_2$~\cite{Biondo2022}; analysis of spatially-average transport models~\cite{Alves2023} and comparison between 1D radial and 0D global models\cite{Viegas2023}. These examples demonstrate that the formulation adopted in LoKI-GM can be adapted to 0D descriptions beyond the strict concept of a global model. In addition, LoKI-GM was benchmarked against the Quantemol Global Model for the pre-assembled oxygen chemistry, available in the Quantemol database~\cite{Tennyson2022}.

The organization of this paper is as follows. Section~\ref{General} provides a general description of the problem and introduces the global model employed. Section~\ref{Numerical} describes the workflow of the code for active plasmas and afterglows, outlines the general input/output parameters, highlights the main differences between the present formulation and those commonly adopted in other global models, and discusses the key aspects of the numerical solution. Section~\ref{Results} presents representative simulation results obtained under different operating conditions, emphasizing the main features of the code and its flexibility for plasma chemistry studies. Finally, Section~\ref{Final} summarizes the main conclusions.

\section{General formulation}
\label{General}
LoKI-GM provides a self-consistent description of the chemical kinetics and transport properties of a plasma or the post-discharge of a plasma with electron density $n_e$, created from a general mixture of atomic/molecular gases at pressure $p$, temperature $T_g$ and input flow $Q_{\rm sccm}$ (in sccm). Each gas $k$ in the mixture has a number density $N_k$, such that $\sum_k N_k = N$, and a corresponding fraction $\chi_k \equiv N_k / N$, such that $\sum_k \chi_k = 1$.  
LoKI-GM uses SI base units for all physical quantities, except the energies that are expressed in eV (1 eV = $1.6 \times 10^{-19}$ J), the reduced electric field that is expressed in Td (1 Td = $10^{-21}$ V m$^2$) and the gas flow that is expressed in sccm (standard cubic centimeters per minute).

When describing active plasmas, the excitation assumes an electric field directed along the $z$-axis $\vec{E}(t) = - E(t) \vec{e}_z$, where $E(t)$ is defined as follows: (i) for \textit{steady-state simulations}, in which LoKI-C and LoKI-B are run in a coupled manner, the time-dependent electric field satisfies
\begin{equation}
E(t) = E_p \cos{(\omega t)}
\label{E-field}
\;\;\; ,
\end{equation}
where the oscillation frequency $\omega$ is either $0$, for a DC excitation, or $\omega \gg \nu_c$  ($\nu_c$ is the typical electron-neutral collision frequency), for a HF excitation. In this case, the electric-field amplitude (and any related quantity, such as the discharge current or the discharge power-density) is self-consistently calculated by imposing charge neutrality (see Section~\ref{Numerical}); (ii) for \textit{quasi-stationary simulations}, in which case LoKI-C runs as a standalone tool, the time-dependent electric field $E(t)$ (or any related quantity, such as the discharge current or the discharge power-density) is defined by the user. In this case, LoKI-GM uses precomputed lookup tables of electron macroscopic parameters, at certain electric-field values, which are then used in the chemistry solution. This configuration imposes charge neutrality, and can be applied to pulsed discharges (see Section~\ref{Numerical}). 

The electric field~\eqref{E-field} can also be applied to \textit{post-discharge simulations}. These simulations typically begin with a coupled LoKI-C/LoKI-B steady-state calculation, followed by a standalone LoKI-C simulation performed at $E = 0$ while imposing charge neutrality, in order to describe the time-dependent post-discharge (see Section~\ref{Numerical}). It is also possible to perform only the post-discharge simulation, starting from suitable initial conditions, by running LoKI-C in standalone mode.
LoKI-GM supports also \textit{electron kinetics simulations} (including swarm analysis), in which case LoKI-B runs as a standalone tool for $E/N$ specified as discrete value(s) or as a time-dependent (non-oscillatory, $\omega=0$) function~\cite{Tejero2021}; however, the \textit{full time-dependent} coupling between LoKI-B and LoKI-C is not yet supported.

LoKI-GM handles simulations in \textit{any} atomic/molecular gas mixture, considering collisions with \textit{any} volume/surface target states, featuring \textit{any} type of internal degrees of freedom: electronic, vibrational and rotational. The densities of these states are self-consistently calculated when using LoKI-GM or LoKI-C, and are prescribed by the user when running LoKI-B as standalone tool. In the following, we focus mainly on the Chemistry solver LoKI-C, as details of the Boltzmann solver LoKI-B can be found in previously published works~\cite{Tejero2019,Tejero2021}.

LoKI-GM solves the spatially-averaged rate balance equations for the various gas/plasma/surface states, assuming that the creation/destruction of each state X$_j$ is due to several $r$-reactions (as defined by the kinetic scheme), such as
\[
\sum_l a_{l,r}^{(1)} {\rm X}_l \longrightarrow \sum_{l^\prime} a_{{l^\prime},r}^{(2)} {\rm X}_{l^\prime}
\;\;\; ,
\]
where $a_{l,r}^{(1)}$ and $a_{{l^\prime},r}^{(2)}$ are the stoichiometric coefficients of states
X$_l$ and X$_{l^\prime}$, as they appear on the left- and right-hand sides of reaction $r$, respectively; and is also due to radial and/or axial transport within a generalized cylinder of radius $R$ and length $L$. Note that, when $L \rightarrow \infty$ the plasma container becomes an infinitely long cylinder with radius $R$, in which case the axial transport can be neglected, whereas when $R \rightarrow \infty$ it degenerates into a parallel-plate system with length $L$, equivalent to a slab geometry, in which case the radial transport can be neglected. The particle density of a generic state $j$ of gas $k$ is represented by $n_{k_j}$, such that $\sum_{j \epsilon {\rm childless}} n_{k_j} = N_k$ for a sum over states without internal degrees of freedom; the following populations are also defined: $\xi_{k_j} \equiv n_{k_j} / n_{{\rm parent \, of \,} k_j}$ and $\delta_{k_j} \equiv n_{k_j} / N$. 

Under these conditions, the spatially-averaged rate balance equation of a generic $j$-state writes~\cite{Alves2018}:
\begin{equation}
\frac{\partial n_j}{\partial t} = S_j^{\rm chem} + S_j^{\rm transp} + S_j^{\rm flow}
\;\;\; ,
\label{rate-balance-eq}
\end{equation}
where 
\begin{subequations}
\begin{eqnarray}
S_j^{\rm chem} & = & 
\sum_{r={\rm chem}} 
 \left\{ \left[ a_{jr}^{(2)} - a_{jr}^{(1)} \right] k_{r_j}  \Pi_l n_l^{a_{lr}^{(1)}} \right\}
\label{source-chem}
\\
S_j^{\rm transp} & = & 
\left\{
\begin{array}{ll}
\displaystyle
\sum_{r={\rm transp}} a_{jr}^{(2)} \frac{n_l}{\tau_r} - \frac{n_j}{\tau_{{\rm transp}_j}} \,\,\, , \,\, \mbox{for neutral states}
\\
\\
\displaystyle
- \frac{n_j}{\tau_{{\rm transp}_j}} \,\,\, , \,\, \mbox{for charged species}
\end{array}
\right.
\label{source-transp}
\end{eqnarray}
\begin{eqnarray}
S_j^{\rm flow} = - k_{\rm outflow} n_j 
\label{source-flow}
\end{eqnarray}
\end{subequations}
represent the chemistry (volume and surface), the transport and the flow net-source terms, respectively.

In~\eqref{source-chem}, $n_l$ is the particle density of state $l$ and $k_{r_j}$ is the rate coefficient of the $r$-reaction involving the $j$-state. The quantity $n_l$ is written for all states as a {\it volume} density, and therefore it is corrected by the surface-to-volume factor $A/V = 2 (R+L) / (RL)$ when applied to {\it surface} states: physisorbed/chemisorbed states X$_f$ and X$_s$, respectively, and vacant physical/chemical sites F$_v$ and S$_v$, respectively.

The characteristic transport times $\tau_{{\rm transp}_j}$ in~\eqref{source-transp}
are presented and discussed in section~\ref{General-rateCoeffTransp}.

The outflow rate coefficient $k_{\rm outflow}$ in~\eqref{source-flow}, assumed equal for all volume states, influences the calculation of the gas pressure given by the particle conservation equation
\begin{equation}
\sum_j n_j = N = \frac{p}{k_B T_g}
\;\;\; ,
\label{particle-cons}
\end{equation}
where $k_B$ is the Boltzmann constant. The expression of $k_{\rm outflow}$ depends on the flow model adopted, and leads to different numerical implementations of the chemistry solution (see section~\ref{Numerical})
\begin{eqnarray}
k_{\rm outflow} \simeq 
\left\{
\begin{array}{ll}
\displaystyle
\frac{Q_{\rm sccm}^{\rm outflow}}{V} \frac{10^{-6}}{60} 
\frac{T_g(t)}{273.15} \frac{1.013 \times 10^5}{p(t)}
\\
\;\;\;\;\;\; \mbox{when specifying a numerical value}
\\
\;\;\;\;\;\;\;\; \mbox{for the outflow $Q_{\rm sccm}^{\rm outflow}$ (in sccm)}
\\
\\
\displaystyle
\frac{1}{N(t)} \sum_j \left( S_j^{\rm chem} + S_j^{\rm transp} \right) + 
\frac{1}{T_g(t)} \frac{d T_g(t)}{dt}
\\
\;\;\;\;\;\; \mbox{when adopting an isobaric model that imposes}
\\
\;\;\;\;\;\;\;\; \mbox{the isobaric condition $dp/dt = 0$ to~\eqref{particle-cons} and~\eqref{rate-balance-eq}}
\; .
\end{array}
\right.
\label{kout}
\end{eqnarray}
In these equations, $N(t)$, $p(t)$ and $T_g(t)$ represent, respectively, the time-dependent total density, pressure and temperature of the gas.

\subsection{Rate coefficients for volume and surface reactions}
\label{General-rateCoeffVolSurf}
For volume reactions, $k_{r_j}$ can be defined as~\cite{Tejero2019,Guerra2019}
\begin{equation}
k_{r_j} = 
\left\{
\begin{array}{ll}
\gamma \int_0^\infty u \sigma_{r_j}(u) f(u) du 
\,\,\, , \,\, \mbox{for electron collisions}
\\
\\
\displaystyle
\mbox{user-defined expressions, \,\, for heavy-species collisions}
\\
\\
A_{r_j} \beta_{r_j}
\,\,\, , \,\, \mbox{for radiative transitions}
\;\;\; ,
\end{array}
\label{krj-volume}
\right. 
\end{equation}
where $\gamma = \sqrt{2e/m_e}$ with $e$ and $m_e$ the electron charge and mass, respectively; $u$ is the electron kinetic energy, $f$ is the EEDF (obtained from LoKI-B) and $\sigma_{r_j}$ is the electron scattering cross section for the $r$-reaction with the $j$-target; $A_{r_j}$ is the Einstein coefficient for radiative transition $r$ from the $j$-state; and $\beta_{r_j}$ is the corresponding escape factor describing the radiation imprisonment. Currently, LoKI-GM includes escape factor functions for steady-state simulations in cylindrical geometry. 

The heavy-species collision rate coefficients can have any expression defined by the user. Typical examples are constant values, gas temperature power-laws or Arrhenius-type expressions
\begin{equation}
p_{r_j} T^{q_{r_j}} \exp{\left( \pm \frac{T_{\rm ref}}{T}\right)}
\label{krj-volume-hs}
\;\;\; ,
\end{equation}
where $p_{r_j}$ and $q_{r_j}$ are rate coefficient parameters; $T_{\rm ref}$ is a reference temperature and $T$ is a general temperature (\textit{e.g.} the gas temperature $T_g$ of the electron temperature $T_e$).

For two-body reactions of a microkinetic mesoscopic surface model, $k_{r_j}$ can be defined as~\cite{Guerra2019,Guerra2007,Marinov2017}
\begin{equation}
k_{r_j} = 
\left\{
\begin{array}{ll}
\displaystyle
\frac{k_{r_j}^0 \, \exp{\left( - \frac{E_{r_j}}{k_B T} \right)}}{[{\rm F}] + [{\rm S}]} 
\frac{1}{\tau_{r_j}} \,\,\, , \\
\,\,\,\,\, \mbox{for adsorption-like and surface-diffusion-like reactions}
\\
\\
\displaystyle
\exp{\left(- \frac{E_{{dr}_j}}{k_B T_w} \right)} 
\frac{1}{\tau_{r_j}}
\,\,\, , \,\, \mbox{for desorption reactions}
\;\;\; .
\end{array}
\label{krj-surface}
\right.
\end{equation}
Examples of adsorption-like reactions are the physical adsorption, the chemical adsorption and the Eley-Rideal (E-R) recombination, respectively
\begin{eqnarray*}
{\rm X_j + F_v \longrightarrow X_f}
\\
{\rm X_j + S_v \longrightarrow X_s}
\\
{\rm X_j + Y_s \longrightarrow XY + S_v}
\;\;\; ;
\end{eqnarray*}
examples of  surface-diffusion-like reactions are surface transport and the Langmuir-Hinshelwood (L-H) recombination, respectively
\begin{eqnarray*}
{\rm X_f + S_v \longrightarrow X_s + F_v}
\\
{\rm X_f + Y_s \longrightarrow XY + S_v + F_v}
\;\;\; ;
\end{eqnarray*}
and desorption reactions correspond to
\[
{\rm X_f \longrightarrow X_j + F_v}
\;\;\; .
\]
In~\eqref{krj-surface}, [F] and [S] are the {\it volume-equivalent} total densities of physical sites and chemical sites, respectively; $k_{r_j}^0$ is a dimensionless steric factor; $E_{r_j}$
is the activation energy for adsorption and surface diffusion; $E_{{dr}_j}$ is the activation energy for desorption; $T_w$ is the wall temperature; and $\tau_{r_j}$ are the characteristic times of the different surface mechanisms $r$ involving the $j$-state, defined as
\begin{equation}
\tau^{-1}_{r_j} = 
\left\{
\begin{array}{ll}
\tau^{-1}_{{\rm transp}_j}
\,\,\, , \,\, \mbox{for adsorption-like reactions}
\\
\\
\frac{3}{4} \nu_{D_j} \exp{\left( - \frac{E_{D_j}}{k_B T_w} \right)} 
\,\,\, , \,\, \mbox{for surface-diffusion-like reactions} 
\\
\\
\nu_{{dr}_j} 
\,\,\, , \,\, \mbox{for desorption reactions}
\;\;\; ,
\end{array}
\label{surfacetime}
\right.
\end{equation}
where $\tau^{-1}_{{\rm transp}_j}$ is the characteristic transport frequency of neutrals (see below);  $\nu_{D_j}$ and $\nu_{{dr}_j}$ are the frequencies of surface diffusion and desorption, respectively; $E_{D_j}$ is the energy barrier for surface diffusion; and the (optional) factor $3/4$ is taken only for surface transport reactions as a correction to back diffusion~\cite{Guerra2007}, assuming that the remaining $1/4$ of the species entering the collection zone can reach a chemisorption site.

\subsection{Rate coefficients for the transport of species}
\label{General-rateCoeffTransp}
Global models are widely used to describe the chemical reactivity of plasmas at high pressures (e.g., above $\simeq 100$~Pa), where charge separation is typically confined to thin sheath regions near the boundaries and can therefore be neglected. Under these conditions, global models may be applied locally at different spatial positions, effectively reducing to \textit{local non-dimensional models} in which transport phenomena can be entirely neglected. At lower (intermediate) pressures, however, \textit{global zero-dimensional} models must account for the formation of space-charge sheaths and the development of density gradients, incorporating their influence through appropriate averaged transport descriptions. LoKI-GM incorporates rate coefficients derived from multiple formulations to describe the transport of both neutral and charged species.

The characteristic transport time of neutral state $j$, appearing in~\eqref{source-transp} and~\eqref{surfacetime}, can be defined from the {\it heuristic Chantry model} as~\cite{Chantry1987}
\begin{equation}
\tau_{{\rm transp}_j} \equiv
\tau_{{\rm diff}_j} + \tau_{{\rm wall}_j} \simeq
\frac{\Lambda^2}{D_j} + 
\frac{1 - \gamma_j / 2}{\gamma_j \langle v_j \rangle} \frac{2RL}{L+R}
\label{transptime}
\;\;\; ,
\end{equation}
by combining the limiting situations of {\it multi-component diffusion transport}, $\tau_{{\rm diff}_j}$, and {\it wall recombination/deexcitation}, $\tau_{{\rm wall}_j}$, assuming no deposition conditions.
In these equations, $D_j$ is the diffusion coefficient of the $j$-state (see more details below); $\Lambda$ is a characteristic diffusion length, estimated assuming a separation of variables in the axial and radial directions, and further considering zero-density boundary conditions at the discharge walls
\begin{equation}
\Lambda^2 \equiv
\left[ \left( \frac{\pi}{L} \right)^2 + \left( \frac{2.405}{R} \right)^2 \right]^{-1}
\label{difflength}
\;\;\; ;
\end{equation}
$\gamma_j$ is the wall-recombination probability of the $j$-state; and 
\begin{equation}
\langle v_j \rangle = \sqrt{\frac{8 k_B T}{\pi M_j}}
\label{vthermal}
\end{equation}
is the thermal velocity of the $j$-state, with $M_j$ the corresponding mass and $T$ a gas-related temperature, to be defined within the transport/thermal model considered. 
Note that equations~\eqref{transptime}-\eqref{difflength} allow retrieving the limiting cases of transport in pure slab and pure cylindrical geometries, by imposing $R \rightarrow \infty$ and $L \rightarrow \infty$, respectively.
Note further that, in \eqref{surfacetime}, the characteristic transport frequency of neutrals is usually defined in the limiting case of wall recombination as 
$\tau^{-1}_{{\rm transp}_j} \simeq \tau^{-1}_{{\rm wall}_j} \simeq \gamma_j (\langle v_j \rangle/4) (A/V)$,
which can be obtained from the second term in~\eqref{transptime}, by neglecting $\gamma_j/2$.

A more evolved {\it multi-component transport} model for neutrals departs from the general flux boundary-condition~\cite{Chantry1987} 
\begin{subequations}
\begin{equation}
\left[ \frac{\nabla n_j}{n_j} \right]_{\rm wall} =
- \frac{\langle v_j \rangle}{4 D_j} \frac{\gamma_j}{1 - \gamma_j / 2} \equiv
- \lambda_j^{-1}
\label{fluxbc}
\end{equation}
applied to the density profile
\begin{equation}
n_j (r,z) = n_{j0} \, J_0 \left( \frac{r}{\Lambda_{r,j}} \right)  \cos{ \left( \frac{z}{\Lambda_{z,j}} \right) }
\;\;\; ,
\end{equation}
\end{subequations}
where $J_0$ is the $0^{\rm th}$ order Bessel function and $\Lambda_{r,j}$ and $\Lambda_{z,j}$
are diffusion lengths for the $j$-state along the radial and the axial directions, respectively. As a result we obtain ($J_1$ is the $1^{\rm st}$ order Bessel function)
\begin{subequations}
\begin{eqnarray}
z_j \tan{(z_j)} & = & \frac{L}{2 \lambda_j} \;\;\; , \;\;\;
\mbox{\rm with $0 \leq z_j = \frac{L}{2 \Lambda_{z,j}} \leq \frac{\pi}{2}$} 
\\
x_j \frac{J_1 (x_j)}{J_0 (x_j)} & = & \frac{R}{\lambda_j} \;\;\; , \;\;\;
\mbox{\rm with $0 \leq x_j = \frac{R}{\Lambda_{r,j}} \leq 2.405$}
\;\;\; ,
\end{eqnarray}
\end{subequations}
which can be solved numerically, together with~\eqref{fluxbc}, to update the characteristic transport time of neutrals \eqref{transptime}, according to
\begin{subequations}
\begin{eqnarray}
\tau_{{\rm transp}_j} & = & \frac{\Lambda_j^2}{D_j}
\label{transptime2}
\\
\frac{1}{\Lambda_j^2} & \equiv & \frac{1}{\Lambda_{r,j}^2} + \frac{1}{\Lambda_{z,j}^2}
\label{difflength2}
\;\;\; .
\end{eqnarray}
\end{subequations}

The multi-component diffusion coefficients $D_j$ in~\eqref{transptime} and~\eqref{transptime2}, describing the collisional transport of the neutrals, are calculated adopting Wilke's model~\cite{Wilke1950} as the limiting case where each neutral species $j$ diffuses in a mixture of stagnant gases 
\begin{equation}
D_j \simeq \frac{1 - \delta_j}{\sum_{i (i \neq j)} \frac{\delta_i}{D_{ji}}}
\;\;\; ,
\end{equation}
where $D_{ji}$ is the binary diffusion coefficient of the $j$-state in the $i$-state, given by~\cite{Hirschfelder1964,Taylor1993}
\begin{equation}
D_{ji} \equiv \frac{3}{16} 
\frac{\sqrt{4 \pi k_B T_g / (2 M_{ji})}}{N \pi \sigma_{h,ji}^2 \Omega_{ji}^{(1,1)}(T^\star)}
\end{equation}
with
\begin{subequations}
\begin{eqnarray}
M_{ji} & \equiv & \frac{M_j M_i}{M_j + M_i}
\label{LennJones1}
\\
\sigma_{h,ji} & \equiv & \frac{1}{2} (\sigma_{h,j} + \sigma_{h,i})
\label{LennJones2}
\\
T^\star & \equiv & \frac{k_B T_g}{\varepsilon_{ji}}
\label{LennJones3}
\\
\varepsilon_{ji} & \equiv & \sqrt{\varepsilon_j \varepsilon_i}
\label{LennJones4}
\\
\Omega_{ji}^{(1,1)}(T^\star) & \simeq & a_1 (T^\star)^{-a_2} + (T^\star + a_3)^{-a_4}
\label{LennJones-collint}
\;\;\; .
\end{eqnarray}
\end{subequations}
In~\eqref{LennJones1}-\eqref{LennJones4}, ($\sigma_{h,j}$, $\varepsilon_j$) and ($\sigma_{h,i}$, $\varepsilon_i$) are the Lennard-Jones parameters for states $j$ and $i$, respectively; and~\eqref{LennJones-collint} is a fitting formula for the collision integral of the same Lennard-Jones potential, where $a_1 = 1.0548$, $a_2 = 0.15504$, $a_3 = 0.55909$ and $a_4 = 2.1705$. 

Simplified {\it binary diffusion} models can also be considered, neglecting the effects of the multi-component transport and replacing the corresponding diffusion coefficients by binary coefficients ($D_j \simeq D_{j,{\rm parent \, gas \,} k}$), with the characteristic diffusion times given by
\begin{equation}
\tau_{{\rm bin-diff}_j} \simeq \frac{\Lambda^2}{D_{j,{\rm parent \, gas \,} k}}
\;\;\; .
\label{bindifftime}
\end{equation} 
These binary models can adopt several possible frameworks: (i) introducing a wall-recombination/deactivation probability $\gamma_j$ to define $1 / \tau_{{\rm transp}_j} \equiv \gamma_j / \tau_{{\rm bin-diff}_j}$; (ii) assuming a uniform distribution of wall-recombination/deactivation probabilities across the $m$ states considered, $\gamma_j \equiv 1/m $, to define $1 / \tau_{{\rm transp}_j} \equiv \gamma_j / \tau_{{\rm bin-diff}_j}$; (iii) replacing $\tau_{{\rm diff}_j}$ with $\tau_{{\rm bin-diff}_j}$ in the heuristic model~\eqref{transptime}; (iv) replacing $D_j$ with the binary coefficient $D_{j,{\rm parent \, gas \,} k}$ in the multi-component transport model~\eqref{transptime2}-\eqref{difflength2}. 

Note that the previous transport models can be adapted for use in a surface chemistry model, by converting the corresponding rate coefficients into equivalents of surface binary-reactions.

The characteristic transport time of charged species $j$, appearing in~\eqref{source-transp}, can be evaluated using two broad classes of models: ambipolar-based formulations and h-factor transport models. A comprehensive review of these approaches is given by Alves and Tejero-del-Caz~\cite{Alves2023}, and only the essential aspects are summarized below. 

In dense plasmas, characterized by $\lambda_D \ll \Lambda$ (with $\lambda_D$ the Debye length), the transport of charged particles can be described adopting \textit{ambipolar-based} models~\cite{Coche2016,Rogoff1985}, which write
\begin{equation}
\tau_{{\rm transp}_j}^{\rm amb} =
\frac{\Lambda^2}{D_{{\rm amb}_j}} 
\;\;\; ,
\label{transptime-ambi}
\end{equation} 
where $D_{{\rm amb}_j}$ is the so-called ambipolar diffusion coefficient of charge species $j$. This general formulation can be realized in various models, which obtain different expressions for $D_{{\rm amb}_j}$ from the solution of the particle and momentum-transfer fluid equations, subject to suitable boundary conditions (usually zero-density at the walls), where the latter equation corresponds to an ambipolar diffusion flux proportional to the density gradient.

The ambipolar diffusion coefficient can be defined for high/low-pressure regimes, by expressing the results of the unified theory proposed by Self and Ewald~\cite{Self1966} in the convenient form of an abacus for an effective (ambipolar) diffusion coefficient $D_{{\rm eff}_j}$ as a function of pressure, as proposed by Ferreira and Ricard~\cite{Ferreira1983} for a single type of positive ions, and later extended by Coche \textit{et al.}~\cite{Coche2016} to plasmas with several positive ions and a single negative ion with low density.

The ambipolar diffusion theory was generalized by Rogoff~\cite{Rogoff1985}, to account for the presence of negative ions, and the resulting framework was subsequently adopted by several authors, under different approximations. The Quantemol Global Model (QGM)~\cite{QGM} clusters the ion fluxes into two equations for single positive ($+$) and negative ($-$) components. The loss frequency of  positive ions is formulated following GlobalKin~\cite{Schroter2018}, accounting for both ambipolar transport and thermal losses at the wall
\begin{subequations}
\begin{equation}
\tau_{{\rm transp}_j}^{\rm QGM} \equiv  
\tau_{{\rm amb}_j} + \tau_{{\rm wall}_j} \simeq
\frac{V}{A} \frac{\Lambda}{D_{a_+}} + 
\frac{V}{A} \frac{4}{\gamma_j \langle v_j \rangle}
\;\;\; ,
\label{transptime-QGM}
\end{equation}
where in general $\gamma_j = 1$ and 
\begin{equation}
{D_{a_+}} \simeq D_+ \frac{1 + T_r \left( 1 + 2 \alpha \right)}{1 + \alpha T_r}
\;\;\; ,
\label{Da+}
\end{equation}
\end{subequations} 
with $T_r \equiv T_e / T_{+}$, $\alpha \equiv n_e/n_{-}$, $D_+ = \sum _i D_i n_i / \sum_i n_i$, and $D_i$ the free-diffusion coefficient for individual positive ions $i$. In plasmas with moderate to low electronegativity, the ambipolar diffusion coefficients derived by Rogoff~\cite{Rogoff1985} may be used when solving the coupled particle and momentum-transfer fluid equations. This formulation was initially proposed by Ferreira \textit{et al.}~\cite{Ferreira1988}, to describe the radial transport in a DC axially homogeneous positive column with three types of charged particles ($+$, $-$ and electrons), and was later extended by Guerra and Loureiro~\cite{Guerra1999} for application to an oxygen plasma with several positive ions. After a considerable number of approximations, one obtains 
\begin{equation}
\tau_{{\rm transp}_j}^{\rm eigen} \simeq \frac{\Lambda_{\rm eigen}^2}{D_{a_j}}
\;\;\; ,
\label{transptime-eigen}
\end{equation}
where $D_{a_j} \simeq T_r D_j$ and the effective diffusion length $\Lambda_{\rm eigen}$ is calculated as an eigenvalue to the problem.

In low pressure discharges, the transport of charged-particles can be described using scaling laws for the density gradient, that rely on the flux of positive ions at the sheath edge, expressed for a single-ion in an electropositive plasma as
\begin{equation}
\Gamma_j = n_{j_s} u_B
\;\;\; ,
\label{ion-edge-flux}
\end{equation}
where $n_{j_s}$ the positive-ion density at the sheath edge, and $u_B = \sqrt{k_B T_e / M_j}$ is the Bohm speed. This flux can be written as a function of the edge-to-center density ratio (designated \textit{h-factor}), and can then be used in the volume-average ion-particle balance equation to express the corresponding effective loss frequency in terms of the h-factors~\cite{Hurlbatt2017,Godyak1986,Lichtenberg1994,Lee1995}
\begin{equation}
\left(\tau_{{\rm transp}_j}^{\rm h-factor}\right)^{-1} = 2 u_B
\left( \frac{h_{L_j}}{L} + \frac{h_{R_j}}{R} \right)
\;\;\; .
\label{transptime-hfac}
\end{equation}
Here, the h-factors $h_{L_j} \equiv n_{{j_s},{\rm axial}}/n_{e_0}$ and $h_{R_j} \equiv n_{{j_s},{\rm radial}}/n_{e_0}$ are the edge-to-center positive-ion density ratios along the axial and the radial directions, respectively.
The h-factors in \eqref{transptime-hfac} can be estimated by equating the flux of positive ions at the sheath edge \eqref{ion-edge-flux} to its expression obtained from different formulations, for example: (i) drift-dominated models for electropositive plasmas, solved numerically at low-pressure~\cite{Godyak1986,Lichtenberg1994} or analytically for plane-parallel electropositive discharges~\cite{Czarnetzki2022}; (ii)  drift-diffusion models, featuring the transition from electropositive to moderate\cite{Lichtenberg1994,Lee1995} or strong~\cite{Kim2006,Thor2010a,Thor2010b,Chabert2016} electronegative plasmas.  

The models expressed by \eqref{transptime}, \eqref{transptime2}-\eqref{difflength2}, \eqref{bindifftime}, for neutral particles, and by \eqref{transptime-ambi}, \eqref{transptime-QGM}-\eqref{Da+}, \eqref{transptime-eigen}, \eqref{transptime-hfac}, for charged particles, are distributed with LoKI-GM as alternative rate-coefficient functions for the transport of species.

\subsection{Thermal model for the gas/plasma}
\label{General-thermalModel}

The results of the plasma chemistry are significantly dependent on the gas temperature, because it affects several rate coefficients for heavy-species collisions causing changes also in the total gas density, at constant pressure. Consequently, global models are often complemented by the thermal balance equation for the gas/plasma, which self-consistently calculates the {\it average} temperature $T_g$ of the heavy-species in the system, assumed thermalised. 

The thermal model adopted in LoKI-GM solves the thermal balance equations in various regions of a cylindrical discharge, for a distributed source-term within the gas/plasma volume and a localized source-term at the discharge wall, considering two alternative boundary conditions: the wall temperature $T_w$ or the external temperature $T_{\rm ext}$.   

In the gas/plasma system, the model considers a radial conduction-flow that sets a one-dimensional parabolic profile for the gas temperature, between the discharge centre ($r=0$) and a radial position near the lateral wall of the cylindrical discharge tube ($r=R_{nw}$)
\begin{equation}
T_g(r) = T_0 - (T_0 - T_{nw}) \frac{r^2}{R_{nw}^2} \;\;\;,
\mbox{for $0 \leq r \leq R_{nw}$} 
\label{Tg-radial}
\;\;\; ,
\end{equation}
with $T_0 \equiv T_g(0)$ and $T_{nw} \equiv T_g(R_{nw})$, which is used to obtain the spatially-averaged gas/plasma thermal balance equation~\cite{Alves2018,Pintassilgo2014} 
\begin{eqnarray}
\sum_k N_k \frac{C_k}{N_A} \frac{\partial T_g}{\partial t} & =
- \frac{\lambda_{\rm mix} \left[T_g - T_{nw} \right]}{\Lambda^2_{\rm th}} 
\nonumber \\
& +  \left| \frac{\Theta_{\rm el}}{N} \right| N n_e + \sum_{r={\rm vol}} R_r \Delta H_r + f_w \sum_{r={\rm wall}} R_r \Delta H_r
\;\;\; .
\label{eqthermal-gas}
\end{eqnarray}
Here, $T_{nw}$ is the near-wall temperature, $C_k$ is the molar specific-heat at constant pressure (or volume) of the gas $k$; $N_A$ is Avogadro's number; $\Lambda_{\rm th} = R_{nw}/\sqrt{8} \simeq R/\sqrt{8}$ is the thermal diffusion length; $\Theta_{\rm el}/N$ is the net power-density lost by one electron, at unit gas density, in elastic collisions with the heavy species; $R_r$ represents the rate of a volume reaction $r = \mathrm{vol}$ or a wall reaction $r = {\rm wall}$, that involves heavy-species and yields an enthalpy exchange $\Delta H_r$ ($\Delta H_r \gtrless 0$ for exothermic and endothermic reactions, respectively); $f_w$ is a general parameter representing the fraction of enthalpy-exchanged in every wall-recombination reaction that is transferred back to the gas/plasma volume; and $\lambda_{\rm mix}$ is the thermal conductivity of the gas mixture given by~\cite{Hirschfelder1964}
\begin{subequations}
\begin{eqnarray}
\lambda_{\rm mix} & = & \sum_k \lambda_k \left[ 
1 + \sum_{k \ne l} G_{k,l} \frac{\xi_k}{\xi_l} \right]^{-1}
\\ 
G_{k,l} & = & \frac{1.065}{2 \sqrt{2}} \left( 1 + \frac{M_k}{M_l} \right)^{-1/2}
\left[ 1 + \left( \frac{\lambda_k^0}{\lambda_l^0} \right)^{1/2} 
\left( \frac{M_k}{M_l} \right)^{1/4} \right]^2
\\
\lambda_k^0 & \equiv & \frac{\lambda_k}{E_k}
\;\;\; ,
\end{eqnarray}
with $\lambda_k$ the thermal conductivity of the gas $k$ and $E_k$ the corresponding Eucken factor. An improved value for the latter factor is given by~\cite{Hirschfelder1964}
\begin{equation}
E_k = 0.115 + 0.354 \frac{C_{p,k}}{k_B N_A}
\;\;\;
\end{equation}
\end{subequations}
with $C_{p,k}$ the molar specific-heat at constant pressure of the gas $k$. 

The temperature at the plasma centre can be expressed as a function of the average and the near-wall temperatures as follows:
\begin{equation}
T_g \equiv \frac{1}{R_{nw}^2/2} \int_0^{R_{nw}} T_g(r) r dr = \frac{T_0 + T_{nw}}{2} \Longrightarrow 
T_0 = 2 T_g - T_{nw}
\label{Tg-average}
\;\;\; .
\end{equation}

In~\eqref{eqthermal-gas}, the term on the left-hand side represents the intrinsic time-variation of the gas thermal-power; whereas the terms on the right-hand side represent, in order, the power lost in conduction, the power gained in electron-neutral elastic collisions, and the net power gained due to the different volume/wall reactions involving heavy-species. 

In the near-wall region, the thermal balance equation yields the equality between the power per unit area flowing outward the gas/plasma system by conduction, $\Gamma_{\rm cond} (R_{nw})$, and flowing inward the region near the tube wall by convection, $\Gamma_{\rm conv, int}$
\begin{subequations}
\begin{eqnarray}
\left\{
\begin{array}{ll}
\Gamma_{\rm cond} (R_{nw}) \simeq \frac{4 \lambda_{\rm mix}}{R} (T_g - T_{nw})
\\
\\
\Gamma_{\rm conv, int} = h_{\rm int} (T_{nw} - T_w)  
\end{array}
\right.
\\
\;\;\;\;\; \Gamma_{\rm cond} (R_{nw}) = \Gamma_{\rm conv, int} 
\Longrightarrow 
\displaystyle T_{nw} = \frac{\frac{4 \lambda_{\rm mix}}{R h_{\rm int}} \, T_g + T_w}{\frac{4 \lambda_{\rm mix}}{R h_{\rm int}} + 1}
\label{eqthermal-nw}
\;\;\; ,
\end{eqnarray}
\end{subequations}
where $h_{\rm int}$ is the convection coefficient in the inner region near the tube wall.

In the region outside the cylindrical tube, the model solves the thermal balance equation across the tube wall (neglecting the effects of conduction within the dielectric), with a localized source-term at the wall boundary due to the enthalpy-exchange in wall-reactions, and considering an external convection flow $\Gamma_{\rm conv, ext}$
\begin{equation*}
\hspace{-2.5cm}
\left\{
\begin{array}{ll}
\left( \Gamma_{\rm conv, ext} - \Gamma_{\rm conv, int} \right) A = \sum_{r={\rm wall}} R_r \Delta H_r V
\\
\\
\Gamma_{\rm conv, ext} = h_{\rm ext} (T_w - T_{\rm ext})  
\end{array}
\right.
\end{equation*}
\begin{subequations}
\begin{eqnarray}
\hspace{-2cm}
\Longrightarrow T_w = \frac{h_{\rm ext} T_{\rm ext} + h_{\rm int} T_{nw} + \sum_{r={\rm wall}} R_r \Delta H_r (R/2)}{ h_{\rm ext} + h_{\rm int}}
\label{eqthermal-out1}
\\
\hspace{-2cm}
\Longrightarrow
\left\{
\begin{array}{ll}
\displaystyle T_w = \frac{\frac{4 \lambda_{\rm mix}}{R h_{\rm ext}} \, T_g + 
\left( \frac{4 \lambda_{\rm mix}}{R h_{\rm int}} +1 \right) T_{\rm ext} + 
\left( \frac{4 \lambda_{\rm mix}}{R h_{\rm int} h_{\rm ext}} + \frac{1}{h_{\rm ext}} \right) 
\sum_{r={\rm wall}} R_r \Delta H_r (R/2)}
{\frac{4 \lambda_{\rm mix}}{R h_{\rm int}} + 
\frac{4 \lambda_{\rm mix}}{R h_{\rm ext}} + 1}
\\
\\
\displaystyle T_{nw} = \frac{\left( \frac{4 \lambda_{\rm mix}}{R h_{\rm int}} + 
\frac{4 \lambda_{\rm mix}}{R h_{\rm ext}} \right) T_g + 
T_{\rm ext} + 
\frac{1}{h_{\rm ext}} \sum_{r={\rm wall}} R_r \Delta H_r (R/2)}
{\frac{4 \lambda_{\rm mix}}{R h_{\rm int}} + 
\frac{4 \lambda_{\rm mix}}{R h_{\rm ext}} + 1}
\;\;\; ,
\end{array}
\right.
\label{eqthermal-out2}
\end{eqnarray}
\end{subequations}
where~\eqref{eqthermal-out2} were obtained combining~\eqref{eqthermal-nw}-\eqref{eqthermal-out1}, and $h_{\rm ext}$ is the convection coefficient in the region outside the cylindrical wall. As expected, by setting $h_{\rm int} , h_{\rm ext} \gg 1$ in equations~\eqref{eqthermal-nw}-\eqref{eqthermal-out2}, one obtains $T_{nw} \sim T_w \sim T_{\rm ext}$. 

Note that different sets of equations are solved depending on the boundary condition adopted: (i) when imposing the wall temperature $T_w$, the model solves~\eqref{eqthermal-gas} with $0 \leq f_w \leq 1$, using~\eqref{eqthermal-nw} and the boundary condition $T_w$, to obtain the time-dependent solutions $T_g(t)$ and $T_{nw}(t)$ for source terms $\sum_{r={\rm vol}} R_r(t) \Delta H_r$ and $\sum_{r={\rm wall}} R_r(t) \Delta H_r$; (ii) when imposing the external temperature $T_{\rm ext}$, the model solves~\eqref{eqthermal-gas} at $f_w \equiv 0$, using~\eqref{eqthermal-out2} and the boundary condition $T_{\rm ext}$, to obtain the time-dependent solutions $T_g(t)$, $T_{nw}(t)$ and $T_w(t)$ for source terms $\sum_{r={\rm vol}} R_r(t) \Delta H_r$ and $\sum_{r={\rm wall}} R_r(t) \Delta H_r$. Note also that the gas/plasma thermal model is solved within an approximate framework that couples the time-dependent heat conduction equation with steady-state convection equations, while assuming a parabolic gas-temperature profile that is taken to be valid at all times.

\section{Workflow and numerical solution}
\label{Numerical}

As mentioned, LoKI-GM comprises two modules (LoKI-B and LoKI-C) that can run self-consistently coupled or as standalone tools. 

The general input to the code specifies the working conditions, including the applied reduced electric field and frequency, gas pressure and temperature, electron density (or a related quantity such as discharge current or power density), and, when considering gas mixtures, the species fractions and state populations of the different components, both in the gas phase and at surfaces.
For LoKI-B, the input data additionally comprise the population distributions of the electronic, vibrational, and rotational states of the atomic/molecular species considered (determined self-consistently when coupled with LoKI-C), as well as the relevant sets of electron-scattering cross sections obtained from the open-access database LXCat~\cite{LXCat}.
For LoKI-C, the required inputs further include the geometrical dimensions of the plasma reactor, gas inflow/outflow conditions, wall/external temperatures relevant to the gas/plasma thermal model, the surface site densities when a microkinetic mesoscopic surface model is employed, and the rate coefficients for all reactions considered in the kinetic scheme, including the electron rate coefficients evaluated in LoKI-B.

As output, LoKI-B provides the isotropic and anisotropic components of the electron distribution function, the corresponding electron macroscopic parameters (transport parameters and rate coefficients), and the electron power absorbed from the electric field and transferred to the various collisional channels. LoKI-C yields the densities of states and the reaction rates associated with the processes included in the kinetic scheme.
In a coupled configuration, LoKI-B supplies the electron macroscopic parameters to LoKI-C. These parameters are then used to solve the system of nonlinear, spatially-averaged rate balance equations~\eqref{rate-balance-eq}, together with the gas/plasma thermal balance equation~\eqref{eqthermal-gas}, based on the specified kinetic scheme for the gas/plasma system.

The system of equations~\eqref{rate-balance-eq} and~\eqref{eqthermal-gas} is solved using the ordinary differential equation (ODE) solvers available in MATLAB~\cite{odesetmatlab}, which rely on time-dependent stiff integration algorithms suitable for chemically reactive plasmas. The integration of the rate balance equations proceeds: 
(i) from default initial conditions (adopted when the user does not specify alternative values), namely $n_0 = N$, $n_{l \neq 0} = 0$ and $T_{nw} = T_g$ (with $T_w = T_{\rm ext}$ additionally imposed when an external boundary condition is selected in the thermal model); 
(ii) up to the final time(s) defined by the user (\textit{i.e.}, the discharge time and, when applicable, the post-discharge time), using optional solver settings that include a relative tolerance of $5 \times 10^{-4}$ and a maximum step-size of $0.1$. 

The supported input/output formats comprise both graphical user interfaces (GUI) and text-based files in ASCII, JSON~\cite{Bray2017} and HDF5~\cite{HDF5}. Currently, GUI-based input is available only for LoKI-B; however, future developments will extend this capability to LoKI-GM. In the HDF5 format, all output data are stored in a single, structured file. This file can be accessed and post-processed using Python scripts, through one of the available HDF5 interfaces~\cite{py-hdf5}, or by applying appropriate import filters in data analysis tools such as MATLAB or Origin~\cite{origin}.

\begin{figure}[htbp]
	\centering
	\vspace{-0.4cm}
	\resizebox{0.9\textwidth}{!}{%
	\includegraphics{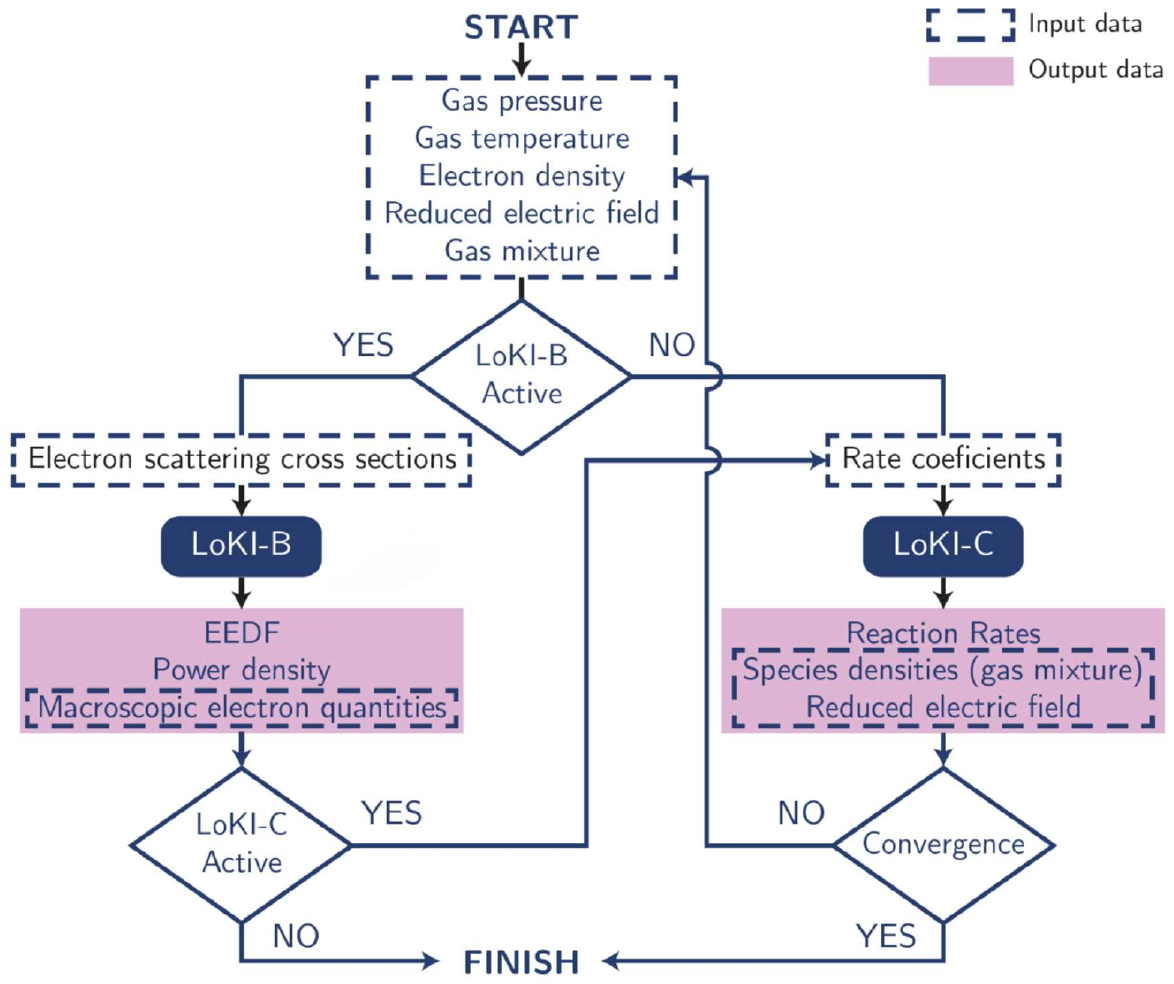}
}
	\vspace{-0.9cm}
	\caption{\label{fig:LoKIWorkflow} General workflow of LoKI-GM. For \textit{steady-state simulations}, in which LoKI-C and LoKI-B are run in a coupled manner, the input electron density can be complemented by values of the discharge current and/or power density. In this case, the \texttt{Convergence} step includes two or three iterative cycles associated with the gas pressure (optional), charge neutrality and the populations of the heavy species relevant to solve the EBE (see text). For \textit{quasi-stationary simulations}, LoKI-B can provide LoKI-C with lookup tables of electron macroscopic parameters tabulated as functions of $E/N$ (for calculations adopting the LFA) and $\varepsilon$ (for calculations adopting the LEA). In this case, LoKI-C runs as a standalone tool (see text). \textit{Post-discharge simulations} start with a coupled LoKI-C/LoKI-B steady-state calculation, followed
by a standalone LoKI-C simulation at $E/N = 0$ to describe the time-dependent post-discharge, using the steady-state values of species densities, electron transport parameters and electron-neutral collisional power transfer (see text). In the latter two cases, no \texttt{Convergence} step is required.}
\end{figure}

Currently, LoKI-GM supports three main types of plasma chemistry simulations, schematically shown in the general workflow of figure~\ref{fig:LoKIWorkflow} and described in the next subsections.

\subsection{Steady-state simulations}
When modeling \textit{active plasmas}, LoKI-GM can implement \textit{steady-state simulations}, in which LoKI-C and LoKI-B are run in a coupled manner. In this framework, LoKI-C updates, at steady-state, the values of the reduced electric field, the populations of the various heavy species and the gas temperature, which are then supplied to LoKI-B for the solution of the electron kinetics.
The numerical workflow involves the following iterative cycles:
\begin{description}
\item [1.] the {\it pressure cycle} (optional), running within LoKI-C to ensure calculations at constant gas-pressure $p$, specified by the user. This cycle verifies, after the full integration of equations~\eqref{rate-balance-eq} and~\eqref{eqthermal-gas}, whether particle conservation~\eqref{particle-cons} is satisfied.
Deviations from condition~\eqref{particle-cons} can arise due to dissociation/recombination/adsorption/desorption processes, which can alter the gas pressure (or the total gas density). If such deviations are detected, the initial value of the total gas density is modified accordingly, and the system of equations is integrated again. This iterative procedure is repeated until equation~\eqref{particle-cons} is satisfied within a user-prescribed tolerance;
\item [2.] the {\it neutrality cycle}, coupling LoKI-C to LoKI-B, to ensure calculations at constant electron density (or any related quantity, such as the discharge current or discharge power density). This cycle verifies, after the full integration of equations~\eqref{rate-balance-eq} and~\eqref{eqthermal-gas}, whether the charge-neutrality condition is satisfied, 
\begin{subequations}
\begin{equation}
\sum_{j_+} n_{j_+} - \sum_{j_-} n_{j_-} = n_e
\;\;\; ,
\label{chargeCons}
\end{equation}
modifying the reduced electric field $E/N$ accordingly, until convergence within a user-prescribed tolerance. The charge-neutrality condition~\eqref{chargeCons} guarantees that, at steady-state, electron production via ionization processes is exactly balanced by their losses due to volume recombination and transport to the walls, where recombination also occurs.

If the calculations are made at constant discharge current $I_{\rm dc}$ (for a cylindrical setup) or at constant discharge power density $dP_{\rm discharge}/dV$, this iterative cycle modifies also the electron density $n_e$ as to satisfy, alternatively,
\begin{eqnarray}
I_{\rm dc} \simeq e n_e (\mu_e N) \left( \frac{E}{N} \right) \pi R^2 
\label{dischargeCurrent}
\\
\frac{dP_{\rm discharge}}{dV} = e \left( \frac{\Theta_{\rm E}}{N} \right) n_e N
\label{dischargePowerDensity}
\;\;\; ,
\end{eqnarray}
where $\mu_e N$ is the reduced electron mobility and $\Theta_{\rm E}/N$ is the power-density gained by the plasma from the applied electric field, per electron at unit gas density, calculated in LoKI-B using~\cite{Tejero2019,Tejero2021}
\begin{equation}
\frac{\Theta_{\rm E}}{N} = \left( \mu_e N \right) \left( \frac{E}{N} \right)^2 
\;\;\; .
\label{theta-E}
\end{equation}
\end{subequations}
Calculations adopting these configurations are extremely useful for a direct comparison between simulations and experiments, where $n_e$ is seldom available. For this reason, the power density $dP_{\rm discharge}/dV$ is expressed in W~m$^{-3}$ (SI units), whereas $\Theta_{\rm E}/N$ is given in eV~s$^{-1}$~m$^{-3}$ as all other energy-related quantities in LoKI-GM; 
\item [3.] the {\it global cycle}, coupling LoKI-C to LoKI-B, to converge over the populations of the heavy species considered in the solution of the EBE. This iterative procedure is repeated until convergence, within a user-prescribed tolerance.
\end{description}
The convergence criterion for the iterative cycles is defined by requiring that certain relative errors fall below prescribed tolerances. For cycles 1 and 2, this corresponds to satisfying conditions~\eqref{particle-cons} and~\eqref{chargeCons}, respectively; for cycle 3, it corresponds to the relative differences in the electron swarm parameters between two consecutive iterations. Iterative cycles 1 and 2 employ the same algorithm, based on a combination of numerical methods (bisection, Newton–Raphson, and spline interpolation), with a default tolerance of $5 \times 10^{-4}$, which is also adopted for iterative cycle 3.

\begin{figure}[htbp]
	\centering
	\vspace{-0.4cm}
	\resizebox{0.9\textwidth}{!}{%
	\includegraphics{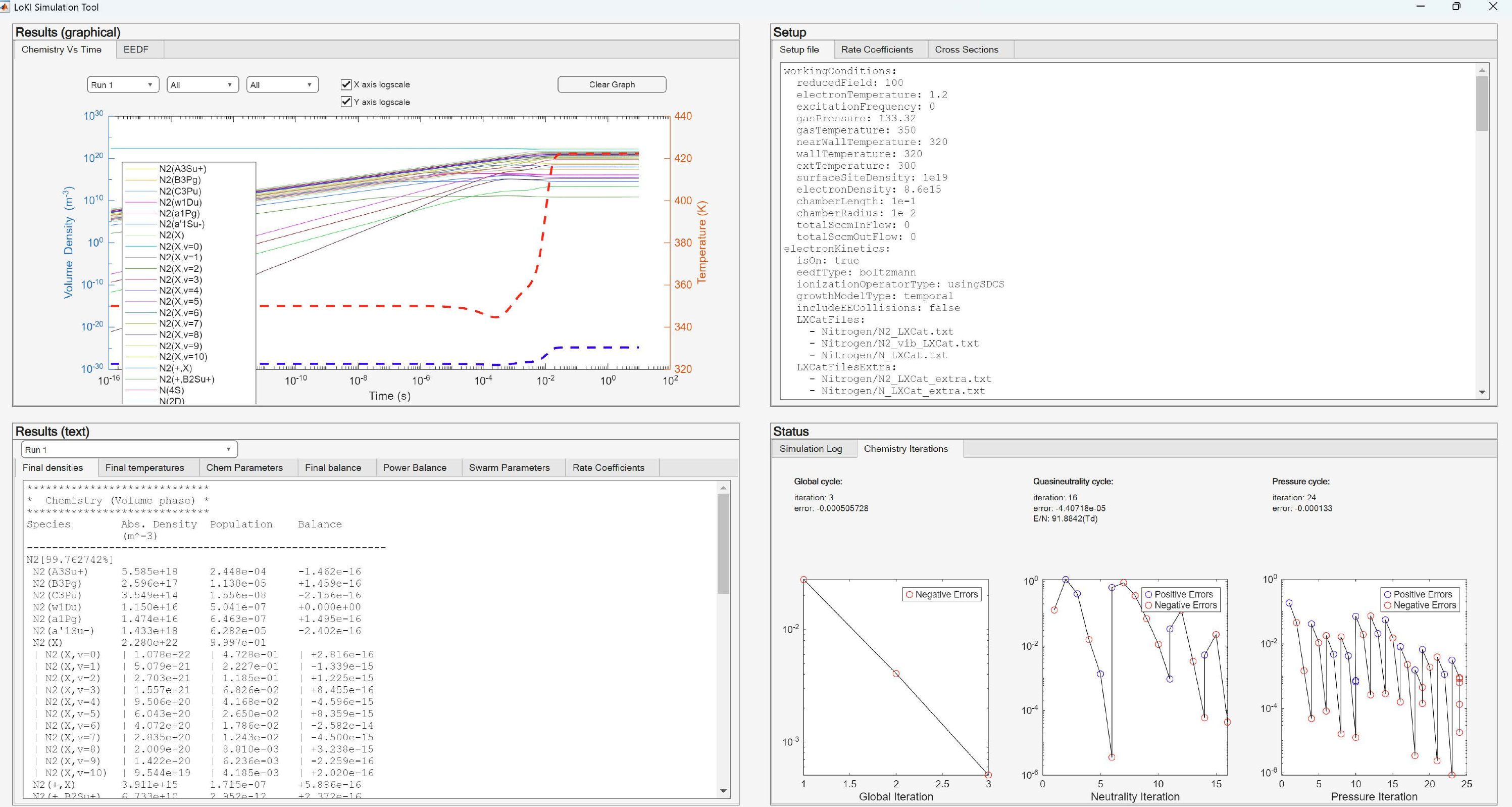}
}
	\vspace{-0.1cm}
	\caption{\label{fig:lokigm-GUI-output} Graphical user interface output of LoKI-GM, for a steady-state simulation in a pure nitrogen DC cylindrical glow discharge ($R = 10^{-2}$~m and $L = 10^{-1}$~m), at $p=133$~Pa and $n_e = 8.6 \times 10^{15}$~m$^{-3}$. The panels are as follows: upper-left, the results for the densities of species (solid curves), the gas temperature (red dashed curve), and the near-wall temperature (violet dashed curve), as a function of the calculation time; upper-right, the summary of the input data; lower-right, the graphs for the three convergence cycles, from right to left, for the pressure, the neutrality, and the global cycle; lower-left, the results for the densities of species in text format.}
\end{figure}

Figure~\ref{fig:lokigm-GUI-output} depicts the graphical user interface output for a steady-state simulation in a pure nitrogen DC cylindrical glow discharge ($R = 10^{-2}$~m and $L = 10^{-1}$~m), at $p=133$~Pa, $n_e = 8.6 \times 10^{15}$~m$^{-3}$ and a wall boundary condition $T_w = 320$~K.
The simulation considered the initial values $E/N=100$~Td, $T_g = 350$~K and $T_{nw} = 320$~K (see upper-right panel in Fig.~\ref{fig:lokigm-GUI-output}), obtaining at steady-state the self-consistent values $E/N \simeq 91.9$~Td (indicated above the middle graph, in the lower-right panel), $T_g \simeq 422$~K and $T_{nw} \simeq 330$~K (see red and violet dashed curves of the plot in the upper-left panel). Additionally, the simulation results yielded $n_{\rm N_2^+} \simeq 3.9 \times 10^{15}$~m$^{-3}$, $n_{\rm N_4^+} \simeq 3.2 \times 10^{15}$~m$^{-3}$, $n_{\rm N_3^+} \simeq 1.5 \times 10^{15}$~m$^{-3}$ and $n_{\rm N^+} \simeq 2.3 \times 10^{13}$~m$^{-3}$, thus satisfying the charge-neutrality condition~\eqref{chargeCons} for $n_e = 8.6 \times 10^{15}$~m$^{-3}$. 

As mentioned, the pressure cycle is optional, since its activation depends on the outflow model adopted (see section~\ref{General}). When $k_{\rm outflow}$ is computed by specifying a numerical value for the outflow $Q_{\rm sccm}^{\rm outflow}$ (top equation of~\eqref{kout}), the steady-state pressure is obtained by iterating on the gas density within the pressure cycle; alternatively, when an isobaric model is adopted (bottom equation of~\eqref{kout}), the steady-state pressure is maintained self-consistently during the solution of the rate-balance equations~\eqref{rate-balance-eq}.

The neutrality cycle can also be deactivated (\textit{e.g.}, by setting to $1$ the number of iterations), in which case LoKI-B and LoKI-C run for the values of $E/N$ and $n_e$ specified by the user. Naturally, the charge conservation is not satisfied in this case, but the configuration can be of interest to assess the populations and the creation/destruction mechanisms of heavy species for certain excitation conditions.

\subsection{Quasi-stationary simulations}
When modeling \textit{active plasmas}, LoKI-GM can also implement \textit{quasi-stationary simulations}, in which LoKI-C runs as a standalone tool. In this framework, the code calculates the time-dependent populations of the various heavy species, while accounting for the prescribed temporal evolution of the electric field (or any related quantity, such as the discharge current or the discharge power-density), specified by the user. The calculations impose the charge-neutrality condition~\eqref{chargeCons} at each instant in time. 
The resulting electron density $n_e(t)$ can then be used into~\eqref{dischargeCurrent} or~\eqref{dischargePowerDensity} to determine the time-dependent reduced electric field $(E/N)(t)$ from the prescribed temporal profiles of the discharge current $I(t)$ or the discharge power density $(dP_{\rm discharge}/dV)(t)$. 

Within \textit{quasi-stationary simulations}, precomputed lookup tables of electron macroscopic parameters, tabulated as functions of $E/N$ and $\varepsilon$, are used in LoKI-GM to interpolate, at selected time instants, the corresponding electron transport parameters, rate coefficients and power transfer required by the chemistry solver. The interpolation procedure can follow one of two approaches: (i) it may use directly the values of the reduced electric field $(E/N)(t)$, obtained from the input data, which corresponds to the LFA; (ii) it may use the electron mean energy $\varepsilon(t)$, computed self-consistently within the chemistry module, which corresponds to the LEA. In the latter case, the set of rate balance equations is solved simultaneously with the electron mean energy equation
\begin{equation}
\frac{d \varepsilon}{dt} = N \left[ \frac{\Theta_E}{N}(t) + 
\left( \frac{\Theta_{\rm coll}^{\rm gain}}{N} \right)(t) 
+ \left( \frac{\Theta_{\rm coll}^{\rm loss}}{N} \right)(t) \right]
\;\;\; ,
\label{mean-energy-eq}
\end{equation}
where $\Theta_E/N$ is calculated using~\eqref{theta-E}, and $\left( \Theta_{\rm coll}^{\rm gain}/N \right)(t)$ and $\left( \Theta_{\rm coll}^{\rm loss}/N \right)(t)$ are the electron 
power densities lost/gained in inelastic/superelastic collisions, respectively, per electron at unit gas density. These quantities are obtained from the precomputed lookup tables employed by LoKI-GM. Note that the gain term is positively defined, whereas the loss term is negatively defined.

\subsection{Post-discharge simulations}
In \textit{post-discharge simulations}, LoKI-GM first computes the corresponding steady-state solution, determining the species densities and the gas temperature. The code then sets the inelastic/superelastic electron rate coefficients to zero (corresponding to a condition with $E/N = 0$), and integrates in time the system of equations~\eqref{rate-balance-eq} and~\eqref{eqthermal-gas}, under the following assumptions: (i) the steady-state species densities and gas temperature are used as initial conditions; (ii) the electron transport parameters and the electron power exchanged in collisions are kept constant, fixed at their steady-state values; (iii) the electron temperature is imposed equal to the gas temperature, consequently, both become time-dependent; (iv) quasi-neutrality is imposed through condition~\eqref{chargeCons}.

\subsection{Closure conditions}
When coupling electron kinetics with chemical kinetics, the closure conditions adopted in LoKI-GM differ somewhat from those commonly employed in other global models~\cite{Alves2018}.
First, LoKI-GM does not explicitly solve the electron particle balance equation. Instead, it imposes charge neutrality, determining the electron density from the corresponding ion balance equations.
Second, LoKI-GM does not directly integrate an electron energy balance equation. For consistency, it relies on the solution of the homogeneous EBE to obtain information on the electron power distribution.
In general, the electron power balance equation can be expressed as~\cite{Alves2018}
\begin{subequations}
\begin{equation}
\frac{d (\varepsilon n_e)}{dt} = \Theta_{\rm growth} n_e + \Theta_E n_e - \Theta_{\rm transp} n_e + \Theta_{\rm coll}^{\rm gain} n_e + \Theta_{\rm coll}^{\rm loss} n_e
\;\;\; ,
\label{rate-balance-epower-eq}
\end{equation}
where the terms on the right-hand side represent, in order: the variation of power density due to non-conservative processes leading to net electron density growth; the power density gained from the applied electric field; the power density lost through transport; and the power density gained/lost in collisional processes. 
Equation~\eqref{rate-balance-epower-eq} may be rewritten as 
\begin{equation}
\frac{d \varepsilon}{dt} = \Theta_E - \Theta_{\rm transp} + \Theta_{\rm coll}^{\rm gain} + \Theta_{\rm coll}^{\rm loss}
\;\;\; ,
\label{rate-balance-epower-eq2}
\end{equation}
\end{subequations}
since $\left( d n_e/dt \right) \varepsilon \equiv \Theta_{\rm growth} n_e$.  Note that, when transport losses are neglected ($\Theta_{\rm transp} \simeq 0$), this expression reduces to~\eqref{mean-energy-eq}, which in fact can be derived by integrating the two-term homogeneous EBE.

\section{Showcase of simulation results}
\label{Results}

This section showcases representative simulation results obtained with LoKI-GM under different operating conditions. In addition to illustrating the main capabilities of the code and its flexibility for plasma chemistry studies, the section also includes and discusses previously unpublished results.

Moreover, the simulation results presented below can be reproduced using the open-source LoKI-GM framework~\cite{LoKISuiteGit}, together with the kinetic schemes and the elementary datasets distributed with the code for oxygen, carbon dioxide and nitrogen plasmas. This approach is consistent with the recent perspective discussed in \textit{On the practical side: why ready-to-use kinetic mechanisms are not available nowadays in the field of low temperature plasma science?}~\cite{Stari2026}, emphasizing  that ``At present, the low temperature plasma community lacks recommended mechanisms, developed and validated to the same level of detail as combustion mechanisms. Driven by the increasing cross-disciplinarity of plasma physics, the community should aim to invest into improved intercommunication. The goal should be to produce these comprehensive and accurate mechanisms, both to push the field forward but also to lower the barrier to entry into plasma physics for scientists and engineers of other fields.''

\subsection{Simulations in oxygen plasmas}
\label{showCase:O2}

Oxygen discharges have been extensively investigated by our group in previous studies~\cite{Dias2023,Viegas2024,Viegas2025,Annusova2018,Marinov2013,Tennyson2022}. In particular, by validating LoKI-GM simulations against measurements obtained from benchmark experiments, we defined a \textit{reaction mechanism}~\cite{Stari2026,Adamovich2017} for DC glow discharges in oxygen, at gas pressures of $0.2-10$~Torr, currents of $10-40$~mA and gas flow rates varying linearly from 2 sccm at 0.2 Torr to 10 sccm at 10 Torr~\cite{Dias2023}. The experimental setup described in~\cite{Booth2019,Booth2020,Booth2022} employed a Pyrex tube with an internal radius of $R = 1$~cm and a length of $L = 52.5$~cm. The wall temperature was maintained at a constant value of $T_w = 323.15$~K using a thermostatic bath. The validation included comparisons not only of the densities of the main species, O$_2$(X), O$_2$(a), O$_2$(b) and O($^3$P), but also of the reduced electric field and the gas temperature. In the present work, we focus on aspects and operating conditions that were not addressed in~\cite{Dias2023}.

\begin{figure}[htbp]
	\centering
	\vspace{-0.4cm}
	\resizebox{0.8\textwidth}{!}{%
	\includegraphics{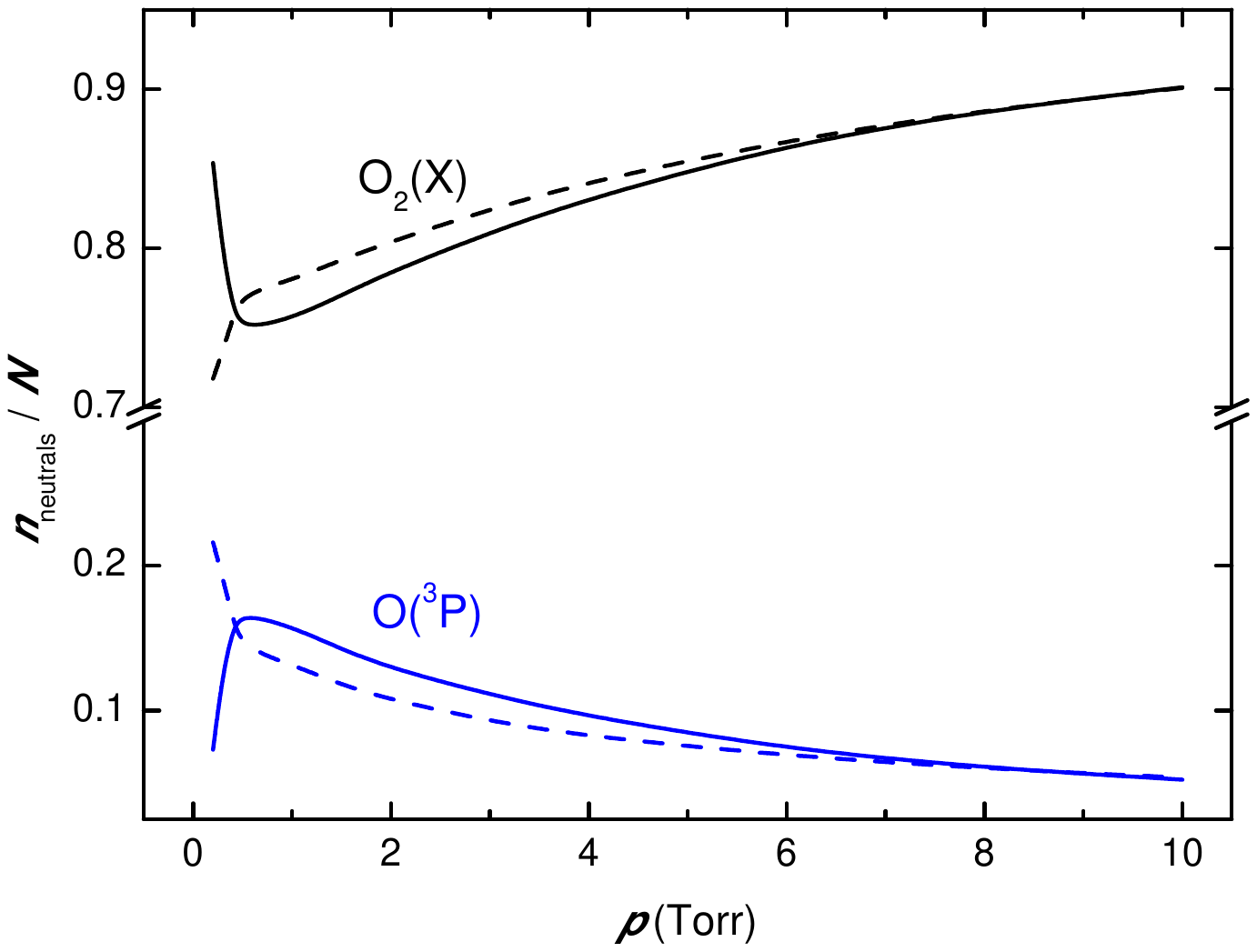}
}
	\vspace{-0.3cm}
	\caption{\label{figO2_constGamma} Relative densities (normalized to the total gas density) of O$_2$(X) (black curves) and O($^3$P) (blue), as a function of pressure, obtained from LoKI-GM simulations of oxygen plasmas sustained by a cylindrical DC discharge ($R = 1$~cm and $L =52.5$~cm), at $I_{\rm dc} = 30$~mA, $T_w = 323.15$~K and with a gas flow rate varying linearly from 2 sccm at 0.2 Torr to 10 sccm at 10 Torr. The simulations were performed using either the pressure- and current-dependent wall-recombination probabilities for O($^3$P) shown in figure 2 of~\cite{Dias2023} (solid curves) or a constant value of $\gamma[{\rm O}(^3{\rm P})] = 2\times10^{-3}$ (dashed).}
\end{figure}

Our previous studies have already demonstrated the critical role played by the wall-recombination of O($^3$P) in the oxygen plasma kinetics~\cite{Dias2023}. This mechanism constitutes the main loss channel for O($^3$P) atoms and, simultaneously, the main production channel for O$_2$(X). In fact, experimental results can only be accurately reproduced when the measured pressure- and current-dependent wall recombination probabilities for O($^3$P)~\cite{Booth2020} $\gamma[{\rm O}(^3{\rm P})]$ (see figure 2 of~\cite{Dias2023}, obtained at $T_w = 323.15$~K), are employed. This effect is illustrated in figure~\ref{figO2_constGamma}, which presents the relative densities (normalized to the total gas density) of O$_2$(X) and O($^3$P), as a function of pressure, calculated with LoKI-GM for the working conditions described above at $I_{\rm dc} = 30$~mA. The simulations 
were performed using either the \textit{standard} variable values of $\gamma[{\rm O}(^3{\rm P})]$ from~\cite{Dias2023} or a constant value of $\gamma[{\rm O}(^3{\rm P})] = 2 \times 10^{-3}$. The constant value underestimates the wall-recombination probability at low pressures, while approaching the measured values at the pressure increases. Consequently, at low $p$, figure~\ref{figO2_constGamma} shows that simulation results are strongly affected by this modification in $\gamma[{\rm O}(^3{\rm P})]$, leading to an increase in the density of O($^3$P) and a corresponding decrease in the density of O$_2$(X). Note that a model describing the dependencies of $\gamma[{\rm O}(^3{\rm P})]$ with pressure and current is presented in~\cite{Viegas2024}.

As previously mentioned, the validation of the reaction mechanism also included comparisons between self-consistent calculations and measurements of the spatially-averaged gas temperature $T_g$. The simulations adopted a gas-to-wall convection coefficient of $h_{\rm int} = 100$~W~m$^{-2}$~K$^{-1}$, together with a boundary condition imposing a fixed wall temperature $T_w$ (see section~\ref{General-thermalModel}). In addition, the calculations assumed that half of the enthalpy-exchange in every wall-recombination reaction is transferred back to the gas/plasma volume, corresponding to $f_w = 0.5$ (see section~\ref{General-thermalModel}). The influence of $h_{\rm int}$ on the radial parabolic profile and on the spatially-averaged values of $T_g$ is discussed in~\cite{Dias2023}, where the dominant gas-heating mechanisms were also identified: wall recombination of O($^3$P) and V–T energy transfer in collisions involving O-atoms.

\begin{figure}[htbp]
	\centering
	\vspace{-0.4cm}
	\subfigure[\label{figO2_Tg}]{\includegraphics[width=3.3in]{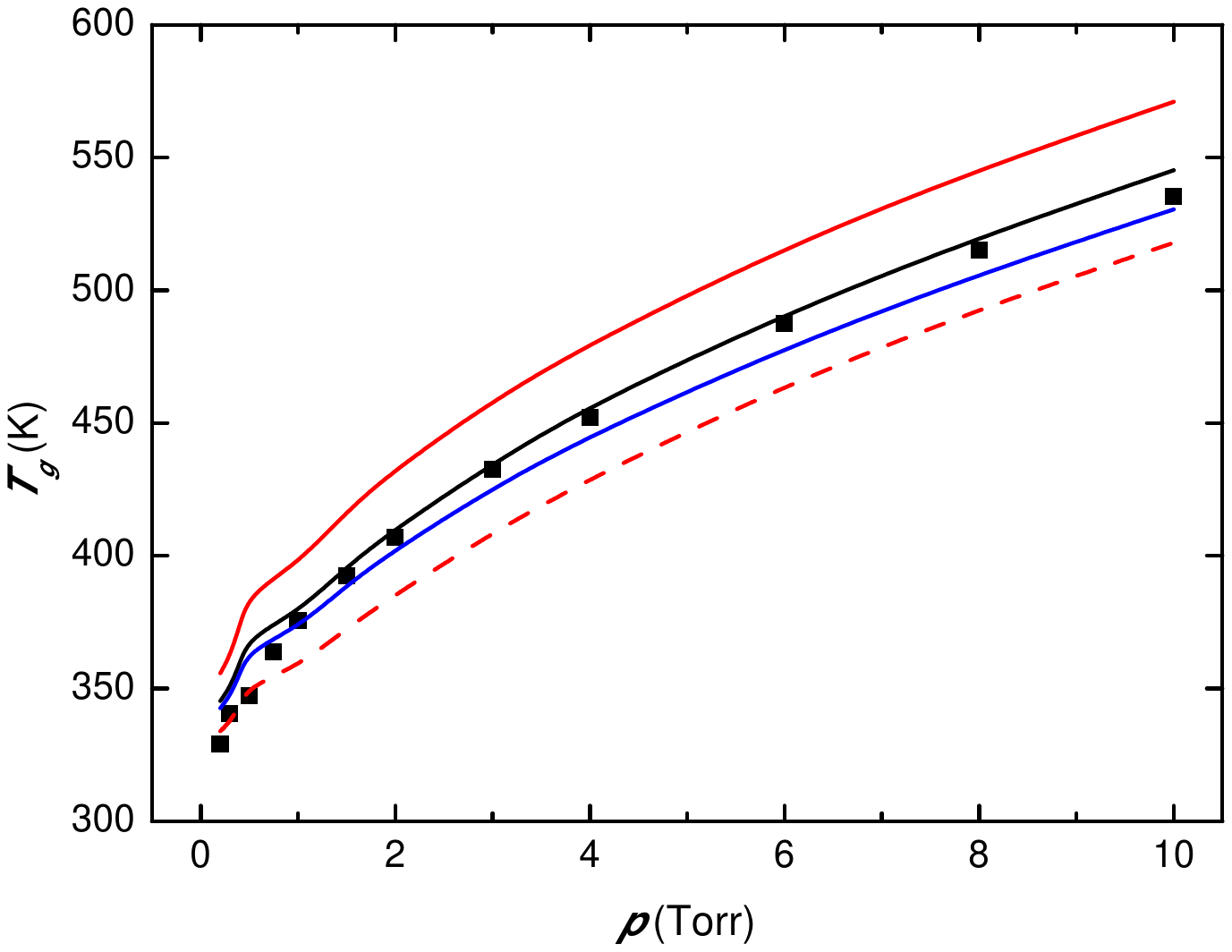}
	\hspace{-1.5cm}}
	\subfigure[\label{figO2_novib-source}]{\includegraphics[width=3.3in]{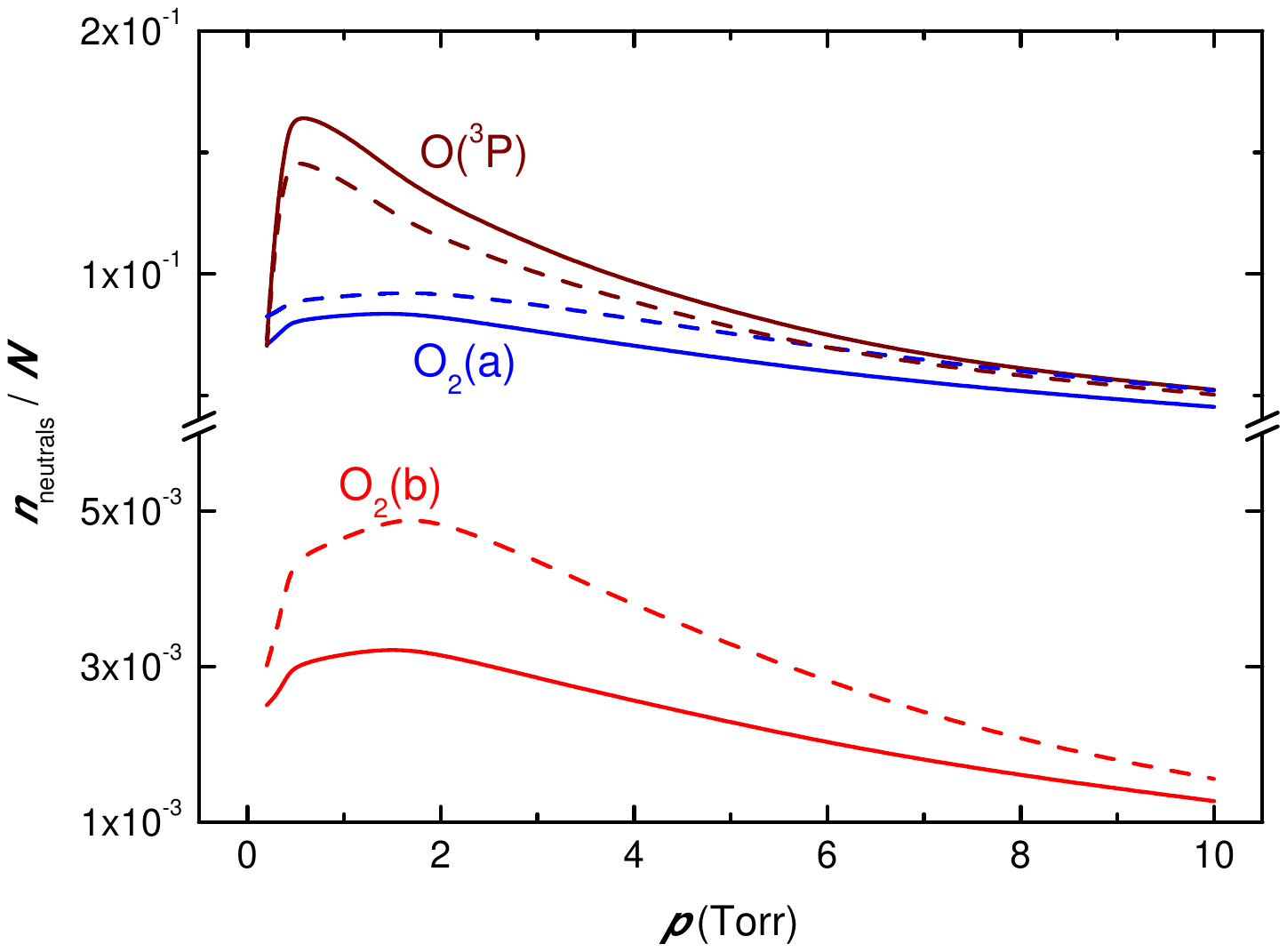}}
	\caption{(a) Gas temperature as a function of pressure, measured~\cite{Booth2019} (points) and calculated (curves) using LoKI-GM in the same conditions as in figure~\ref{figO2_constGamma}, assuming a variable $\gamma[{\rm O}(^3{\rm P})]$. The different curves correspond to the following assumptions: solid black, a fraction $f_w = 0.5$ of enthalpy-exchanged in every wall-recombination reaction that returns to the gas/plasma volume; solid red, $f_w = 1.0$; dashed red, $f_w = 0$; and solid blue, neglecting the contribution of vibrational excitation to the source terms of the gas/plasma thermal balance equation (with $f_w = 0.5$). (b) Relative densities (normalized to the total gas density) of O$_2$(a) (blue curves), O$_2$(b) (red) and O($^3$P) (brown), as a function of pressure, obtained using LoKI-GM in the same conditions as in figure~\ref{figO2_constGamma}, assuming a variable $\gamma[{\rm O}(^3{\rm P})]$. The simulations either include (solid curves) or neglect (dashed) the contribution of vibrational excitation to the source terms of the gas/plasma thermal balance equation.}
	\vspace{-0.3cm}
\end{figure}

Figure~\ref{figO2_Tg} presents $T_g$ vs.~$p$, comparing experimental measurements~\cite{Booth2019} with simulation results obtained using LoKI-GM under the same working conditions, namely the standard configuration previously defined, which adopts $f_w = 0.5$ and includes the contribution of vibrational excitation in the source terms of the gas/plasma thermal balance equation~\eqref{eqthermal-gas}. This figure also shows simulations results obtained by varying the fraction of energy that returns to the volume after a wall-recombination event, considering the limiting cases $f_w=1$ and $f_w=0$, as well as by neglecting the contribution of vibrational excitation in~\eqref{eqthermal-gas}. As expected, the calculated gas temperature decreases when the contribution of gas-heating mechanisms to the thermal balance is reduced.  

Variations in the gas temperature also affect the densities of some oxygen species, as shown in figure~\ref{figO2_novib-source}, which analyzes the calculated relative densities (normalized to the total gas density) of O$_2$(a), O$_2$(b) and O($^3$P), as a function of pressure, for different choices of the source terms in the gas/plasma thermal balance equation. Figure~\ref{figO2_novib-source} shows that neglecting the contribution of vibrational excitation in~\eqref{eqthermal-gas} decreases the relative density of O($^3$P) while increasing the relative densities of O$_2$(a) and O$_2$(b). However, simulation results reveal that the absolute density of O($^3$P) remains essentially unchanged in the two cases considered here. Therefore, the decrease in $n_{\rm O(^3P)}/N$ is solely a consequence of the increase in $N$, resulting from the lower $T_g$. In contrast, the increase in $n_{\rm O_2(a)}/N$ and  $n_{\rm O_2(b)}/N$ indicates that the absolute densities of these excited states also increase when the contribution of vibrational excitation is neglected in~\eqref{eqthermal-gas}.  

\begin{figure}[htbp]
	\centering
	\vspace{-0.4cm}
	\resizebox{0.8\textwidth}{!}{%
	\includegraphics{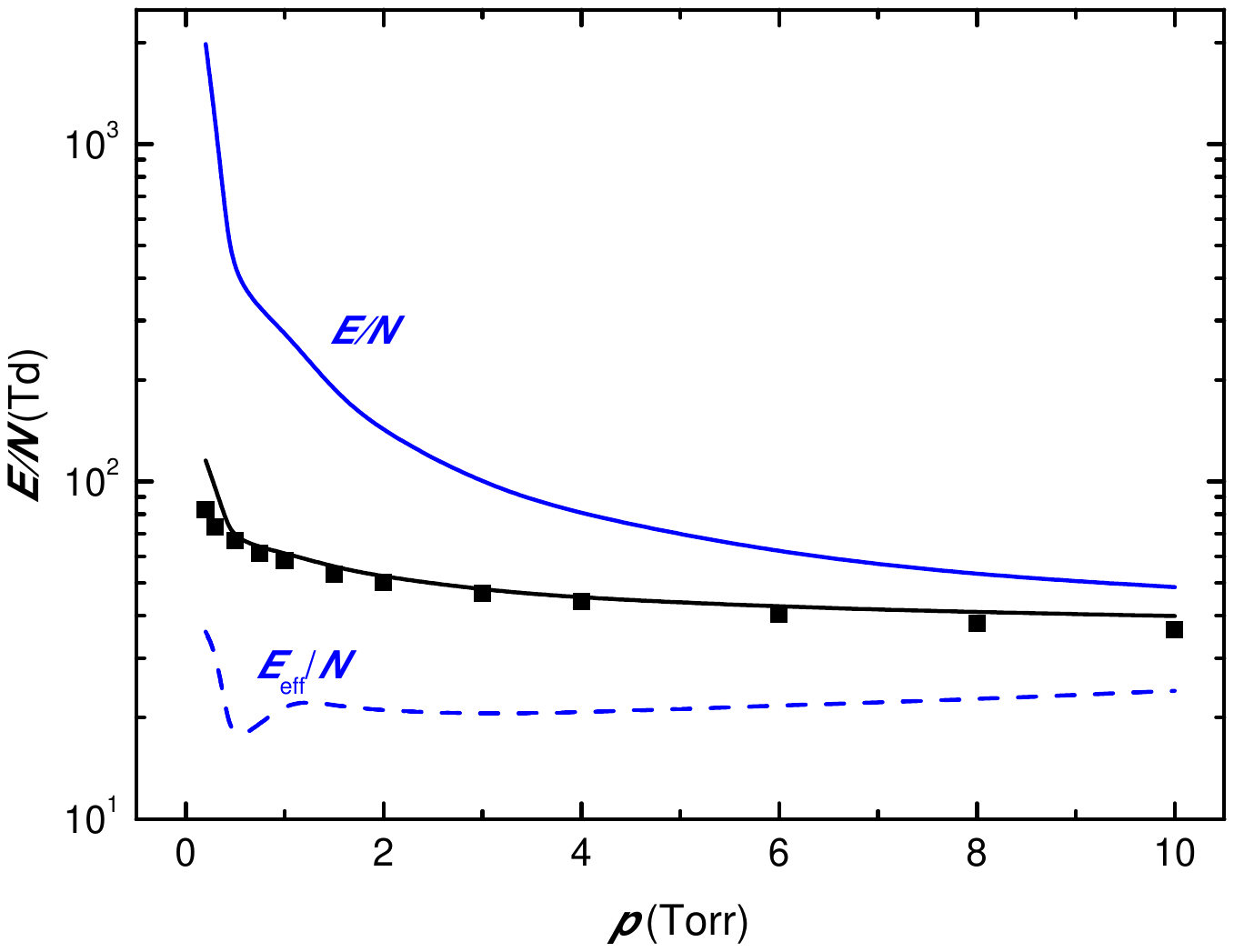}
}
	\vspace{-0.3cm}
	\caption{\label{figO2_dischChar} Discharge characteristics of oxygen plasmas, expressed in terms of the reduced electric field as a function of pressure. The results correspond to the following conditions: measurements~\cite{Booth2019} (points) and LoKI-GM simulations for DC discharges (black curve) performed under the same conditions as in figure~\ref{figO2_constGamma}; LoKI-GM simulations for HF discharges (blue curves) at an excitation frequency of $2.45~$GHz, using the same discharge power-density, reactor geometry, wall temperature and gas flow conditions adopted in figure~\ref{figO2_constGamma}. The solid blue curve denotes $E/N$, whereas the dashed blue curve represents the corresponding effective reduced electric field $E_{\rm eff}/N$.}
\end{figure}

The showcase of simulations for oxygen plasmas concludes with the comparison of discharge characteristics, expressed in terms of the reduced electric field as a function of pressure, as illustrated in figure~\ref{figO2_dischChar}. The figure compares experimental measurements by~\cite{Booth2019} with results obtained using LoKI-GM under the same working conditions previously defined for DC discharges. In addition, figure~\ref{figO2_dischChar} includes simulation results obtained for HF discharges operating at an excitation frequency $\omega / 2 \pi = 2.45$~GHz, considering conditions similar to those adopted for the DC discharges, namely the same discharge power-density, reactor geometry, wall temperature and gas flow rate. 

For HF discharges, figure~\ref{figO2_dischChar} shows the pressure dependence of both the reduced electric field, $E/N$, and the \textit{effective} reduced electric field~\cite{Tejero2019,Alves1992},
\begin{equation}
\frac{E_{\rm eff}}{N} \equiv \frac{E}{N} 
\frac{\langle \nu_c \rangle /N}{\left[ (\langle \nu_c \rangle /N)^2 + (\omega/N)^2 \right]^{1/2}}
\;\;\; ,
\label{EeffoverN}
\end{equation}
which is the quantity employed in the EBE to determine the power effectively transferred to the electrons by the electric field, $\Theta_E/N$. In~\eqref{EeffoverN}, the quantity $\langle \nu_c \rangle /N \simeq 6 \times 10^{-14}$~m$^3$~s$^{-1}$ denotes the average reduced electron-neutral collision frequency, which is approximately constant in the pressure range considered here.
In this case, $\omega / N$ lies in the range $3 \times 10^{-12}$~m$^3$~s$^{-1}$ - $9 \times 10^{-14}$~m$^3$~s$^{-1}$, and therefore one can write approximately 
\begin{equation}
\frac{E_{\rm eff}}{N} \simeq \frac{E}{N} \frac{\langle \nu_c \rangle /N}{\omega/N} 
\propto \frac{E}{N} \times N
\;\;\; .
\label{EeffoverN-bis}
\end{equation}
Equation~\eqref{EeffoverN-bis} can be used to explain the change in the monotonic behavior of $E_{\rm eff}/{N}$ with pressure, observed in figure~\ref{figO2_dischChar}.  At low pressures, the strong decrease of $E/N$ with increasing $p$ leads to a corresponding decrease in $E_{\rm eff}/{N}$. However, as the pressure increases further, the total gas density $N$ also increases (despite the slight rise in $T_g$ observed in the HF case, similar to that shown in figure~\ref{figO2_Tg}), and this increase eventually becomes sufficient to cause an increase in the effective reduced electric field.  

\begin{figure}[htbp]
	\centering
	\vspace{-0.4cm}
	\subfigure[\label{figO2_ne}]{\includegraphics[width=3.3in]{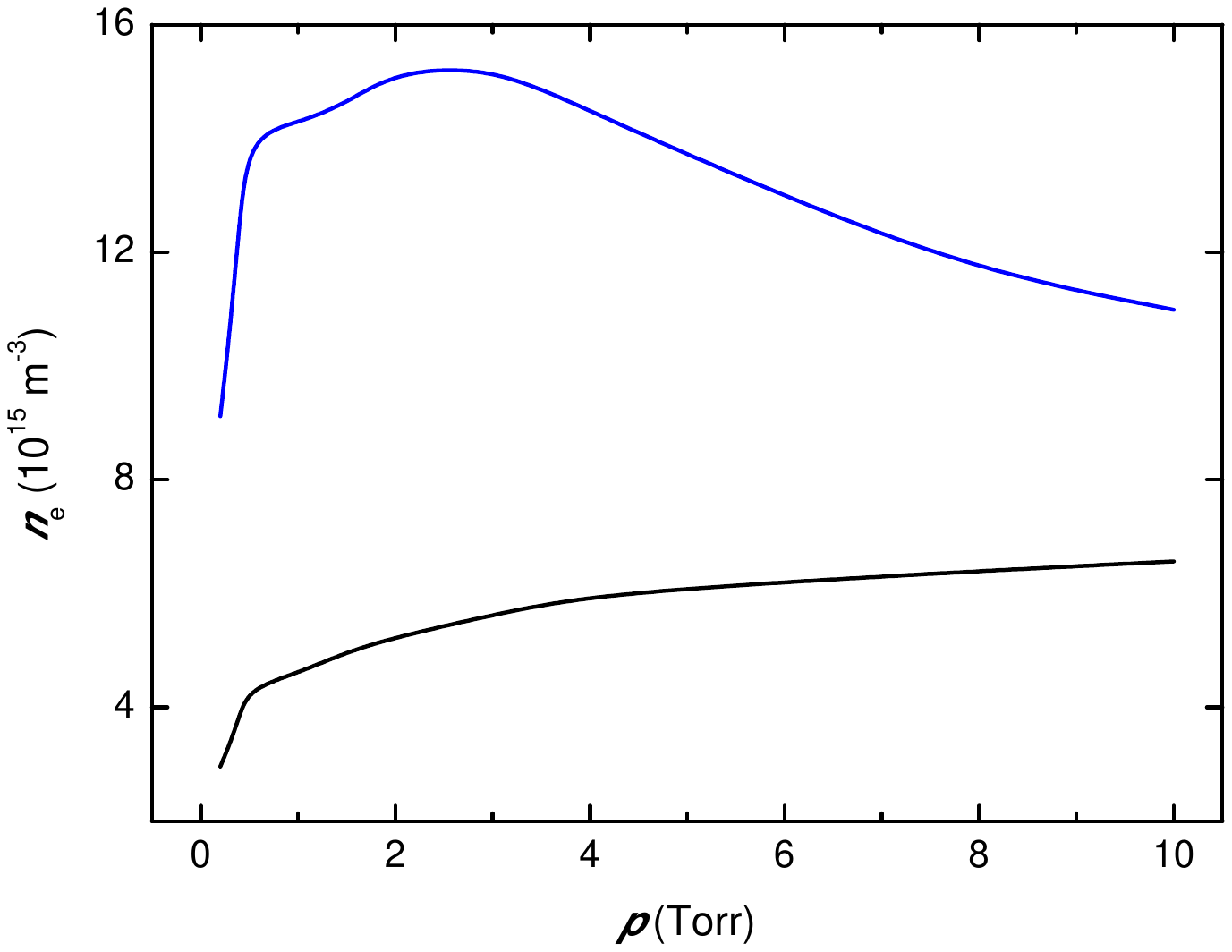}
	\hspace{-1.5cm}}
	\subfigure[\label{figO2_nion}]{\includegraphics[width=3.3in]{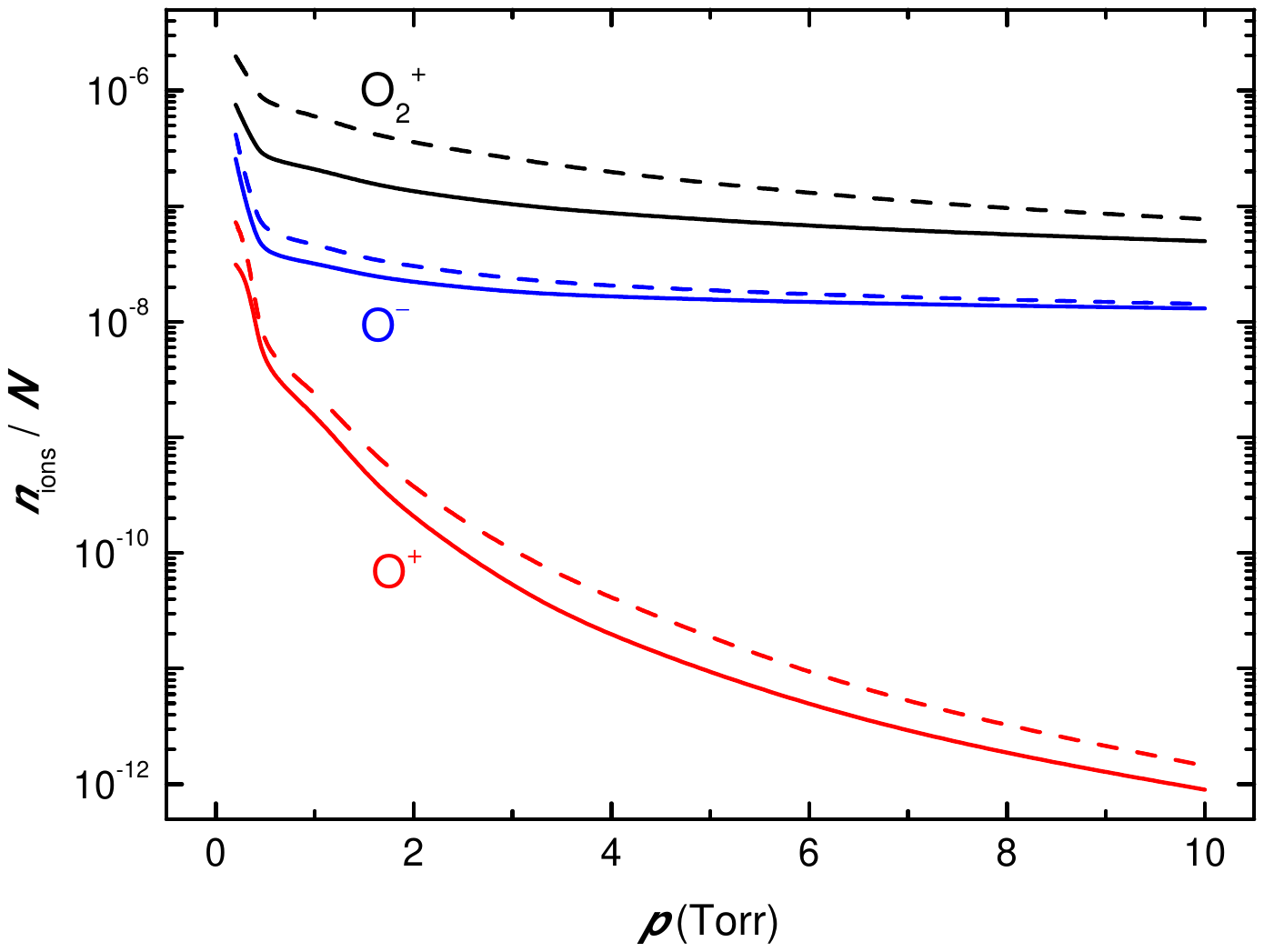}}
	\caption{(a) Electron density as a function of pressure, calculated using LoKI-GM for DC discharges (black curve) and HF discharges (blue), in the same conditions as in figure~\ref{figO2_dischChar}. (b) Relative ion densities (normalized to the total gas density) of O$_2^+$ (black curves), O$^+$ (red) and O$^-$(blue), as a function of pressure, calculated using LoKI-GM for DC discharges (solid curves) and HF discharges (dashed), in the same conditions as in figure~\ref{figO2_dischChar}.}
	\vspace{-0.3cm}
\end{figure}

As mentioned, the HF simulations were performed adopting the discharge power-density $dP/dV$ obtained for the DC case. Under these conditions, the power coupling occurs through a lower effective electric field (see figure~\ref{figO2_dischChar}) and, consequently, through a lower power transferred from the electric field to the electrons, $\Theta_E/N$ (see~\eqref{theta-E}). As a result, a higher electron density is required to sustain the prescribed discharge power-density, (see~\eqref{dischargePowerDensity}), as observed in figure~\ref{figO2_ne}.   As expected, the higher electron density is accompanied by increased densities of the various oxygen ions, O$_2^+$, O$^-$ and O$^+$, as illustrated in figure~\ref{figO2_nion}. The differences between the DC and HF discharge regimes become less pronounced as the pressure increases, owing to the corresponding decrease in the reduced excitation frequency, $\omega/N$ (see~\eqref{EeffoverN}). 

\subsection{Simulations in CO$_2$ plasmas}
\label{showCase:CO2}

CO$_2$ plasmas constitute an important application of global kinetic models because of their relevance to plasma-assisted CO$_2$ conversion and gas processing. In these systems, the gas-flow conditions can strongly influence both the transient plasma dynamics and the resulting chemical composition. The flexibility of LoKI-GM enables the implementation of different gas-flow models, ranging from closed reactors to isobaric systems and continuously fed reactors, thereby allowing the simulation of a wide variety of operating conditions encountered in CO$_2$ plasmas.

\begin{figure}[htbp]
	\centering
	\vspace{-0.4cm}
	\subfigure[\label{figCO2_01}]{\includegraphics[width=3.3in]{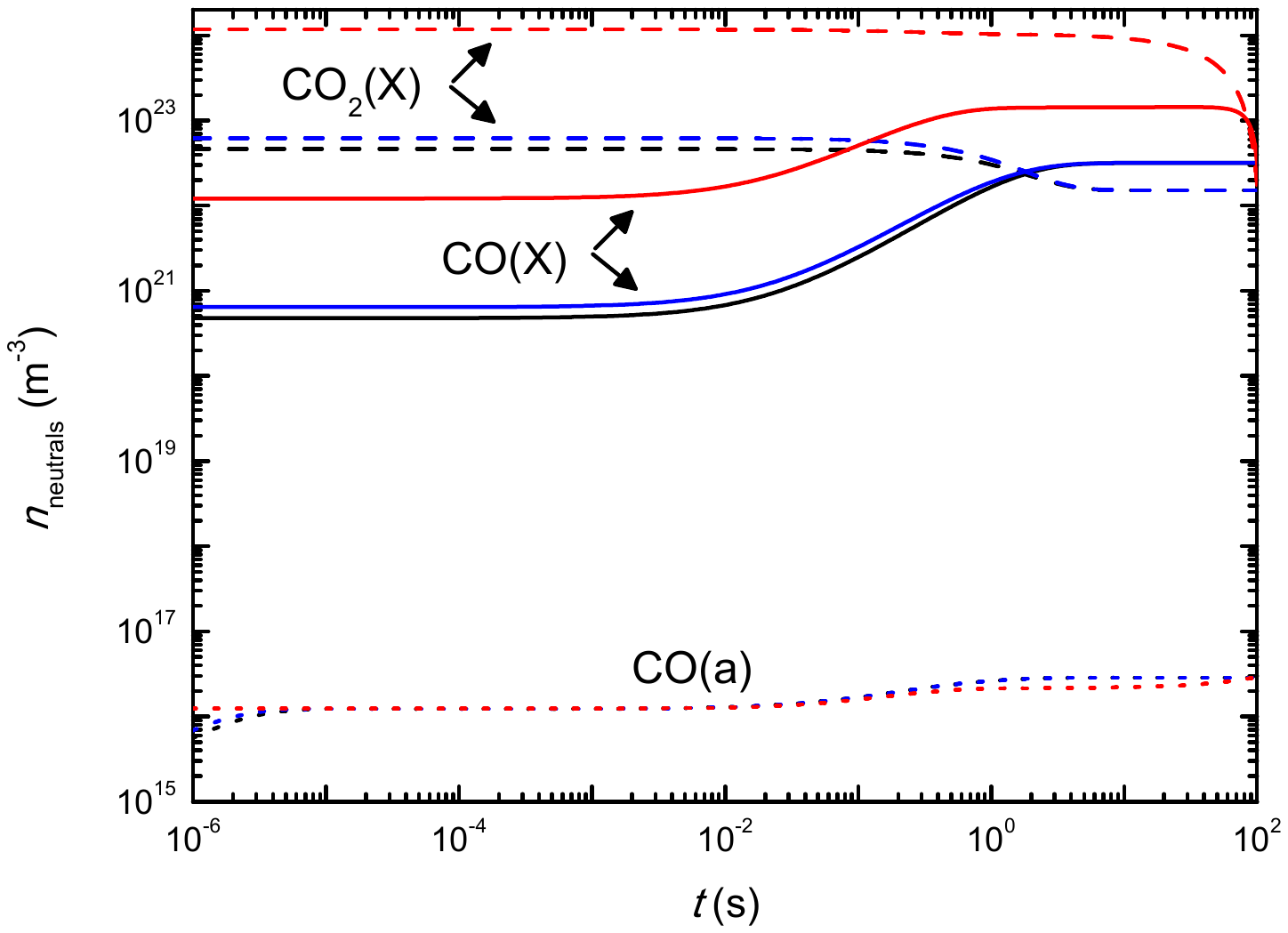}
	\hspace{-1.5cm}}
	\subfigure[\label{figCO2_02}]{\includegraphics[width=3.3in]{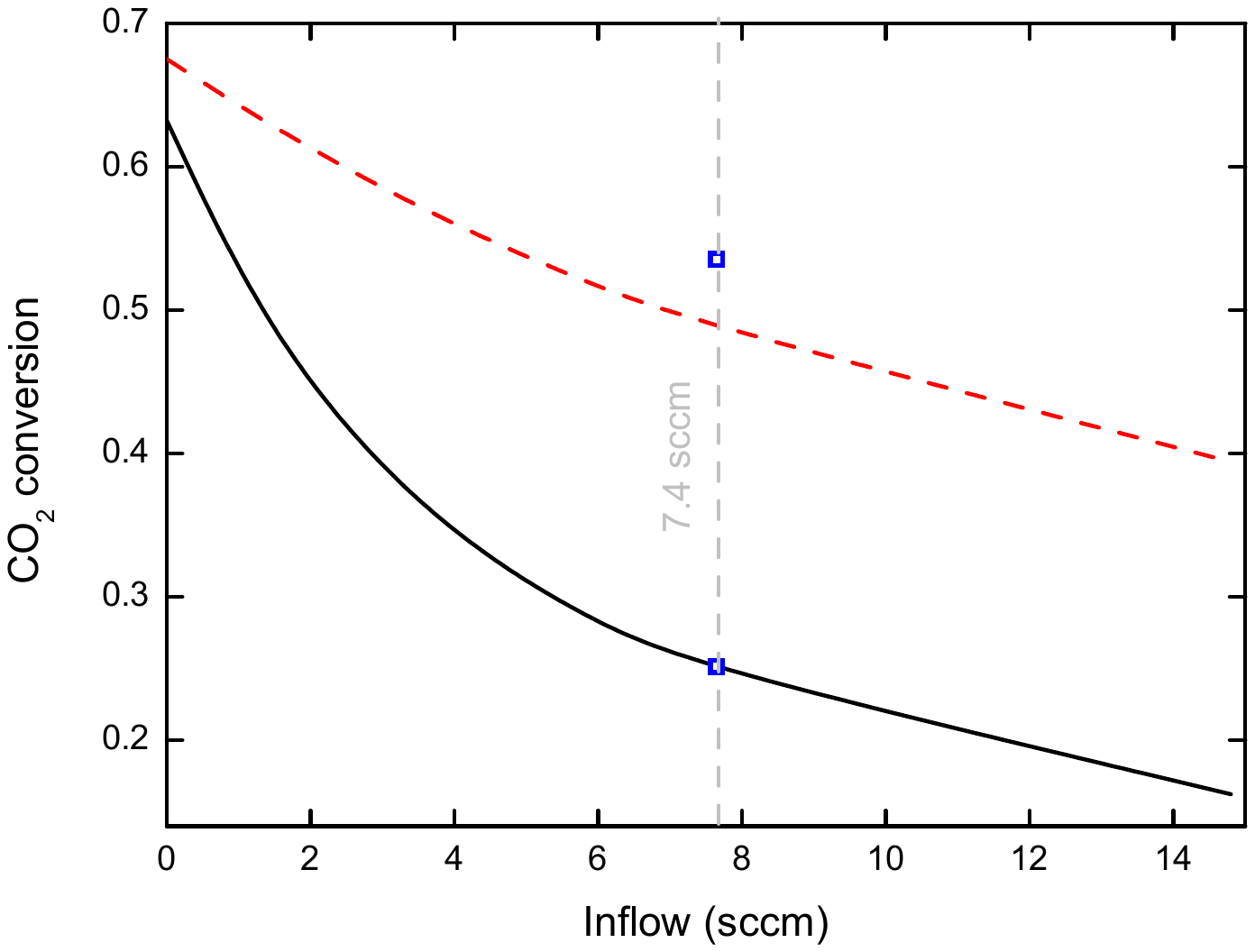}}
	\caption{(a) Temporal evolution of the densities of CO$_2$(X) (dashed), CO(X) (solid), and CO(a) (dotted) in a DC glow discharge sustained in pure CO$_2$ at $p=5$~Torr, $T_g=750$~K, and $I_{\rm dc}=50$~mA, for a cylindrical reactor with radius $R=1$~cm and length $L=67$~cm. The simulations were performed under three gas-flow conditions: (A) a closed reactor with neither inflow nor outflow (black curves); (B) a reactor without inflow using the self-consistent isobaric formulation (blue); and (C) a continuously fed reactor with an inflow of $1$~sccm and an outflow of $7.4$~sccm (red). (b) CO$_2$ conversion, defined as [CO]/([CO]+[CO$_2$]), as a function of the gas flow rate for DC glow discharges sustained in pure CO$_2$. Results are shown for $I_{\rm dc}=10$~mA and $T_g=450$~K (solid black curve), and for $I_{\rm dc}=50$ ~mA and $T_g=750$~K (dashed red curve). The vertical dashed line indicates the benchmark operating flow rate of $7.4$~sccm, while the blue symbols correspond to the experimental conversion measured by Morillo-Candas \textit{et al.}~\cite{MorilloCandas2019} and reported by Liu \textit{et al.}~\cite{Liu2025}.}
	\vspace{-0.3cm}
\end{figure}
Figure~\ref{figCO2_01} compares the temporal evolution of the densities of CO$_2$(X$^1\Sigma_g^+$), CO(X$^1\Sigma^+$) and CO(a$^3\Pi$), calculated for a DC glow discharge sustained in pure CO$_2$ at $p = 5$~Torr, $T_g = 750$~K and $I_{\rm dc} = 50$~mA. The discharge was assumed to be generated in a cylindrical reactor with radius $R = 1$~cm and length $L = 67$~cm, following the experimental conditions reported by Morillo-Candas \textit{et al.}~\cite{MorilloCandas2019}. For simplicity, these states are denoted in Figure~\ref{figCO2_01} as CO$_2$(X), CO(X) and CO(a), respectively. Three different operating conditions were considered: (A) a closed reactor, with neither gas inflow nor outflow; (B) a reactor without inflow, using the self-consistent isobaric formulation described in section~\ref{Numerical}; and (C) a continuously fed reactor with an inflow of $1$~sccm and an outflow of $7.4$~sccm.

For the closed reactor (case A), electron-impact dissociation progressively converts CO$_2$ into CO, leading to a decrease in the density of the parent molecule and a corresponding increase in the concentration of CO(X). The metastable state CO(a) exhibits a similar temporal evolution, reaching densities on the order of 10$^{16}$~m$^{-3}$. The discharge dynamics are governed by the gradual conversion of the initial CO$_2$ into reaction products, with the species densities approaching steady-state after approximately $10$~s$-$$100$~s. Case B yields results that are nearly indistinguishable from those of the closed reactor. This indicates that, under the present operating conditions, the self-consistent isobaric formulation primarily compensates for the pressure variations induced by the chemical kinetics, while having only a minor influence on the dominant reaction pathways and the resulting plasma composition. Consequently, both approaches predict essentially the same temporal evolution for the densities of CO(X), CO$_2$(X) and CO(a). By contrast, the continuously fed reactor (case C) exhibits a markedly different behavior. The continuous supply of fresh CO$_ 2$, together with the simultaneous removal of gas through the outlet at a much higher flow, substantially modifies the balance between chemical production and loss processes. As a result, unlike cases A and B, the discharge does not reach steady-state within the simulated time interval. Instead, the densities of CO$_2$(X), CO(X) and CO(a) continue to evolve, reflecting the persistent competition between plasma-driven conversion and gas renewal.

Figure~\ref{figCO2_02} complements the previous results by presenting the CO$_2$(X) conversion, defined as [CO(X)] / ([CO(X)] + [CO$_2$(X)]), as a function of the gas flow rate. The calculations were performed for two discharge conditions: $I_{\rm dc} = 10$~mA and $T_g = 450$~K, and $I_{\rm dc} = 50$~mA and $T_g = 750$~K. The vertical dashed line indicates the flow rate of $7.4$~sccm, corresponding to the benchmark operating conditions previously validated against the experimental measurements reported by Liu \textit{et al.}~\cite{Liu2025}. As expected, the CO$_2$ conversion decreases with increasing gas flow rate under both discharge conditions. Increasing the flow rate continuously replenishes the reactor with fresh CO$_2$, while simultaneously reducing the residence time of the gas in the reactor, thereby limiting the extent of plasma-induced dissociation. In the limit of very low flow rates, corresponding to conditions approaching those of a closed reactor, the conversion exceeds $60\%$. As the flow rate increases, however, the conversion decreases markedly. The discharge current also has a pronounced effect on the conversion efficiency. For all flow rates considered, the simulations predict substantially higher conversion at $I_{\rm dc} = 50$~mA than at $I_{\rm dc} = 10$~mA. This behavior reflects the more efficient electron-impact dissociation associated with the larger electron density and power deposition at higher discharge currents. For the benchmark flow rate of $7.4$~sccm, the model predicts a conversion of approximately $25\%$ at $I_{\rm dc} = 10$~mA and approximately $50\%$ at $I_{\rm dc} = 50$~mA.

\begin{figure}[htbp]
	\centering
	\vspace{-0.4cm}
	\subfigure[\label{figCO2_03}]{\includegraphics[width=3in]{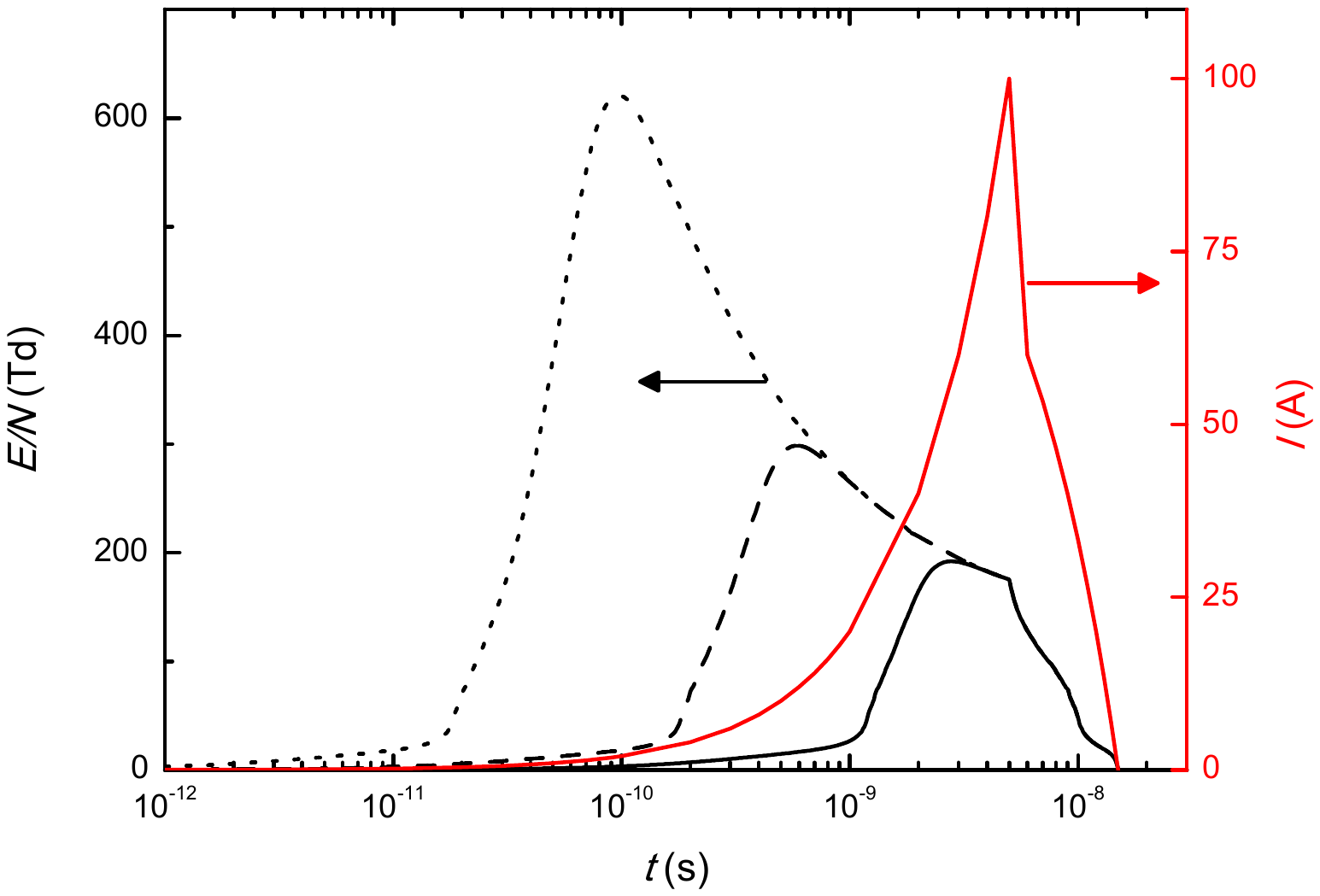}
	\hspace{-1cm}}
	\subfigure[\label{figCO2_04}]{\includegraphics[width=3in]{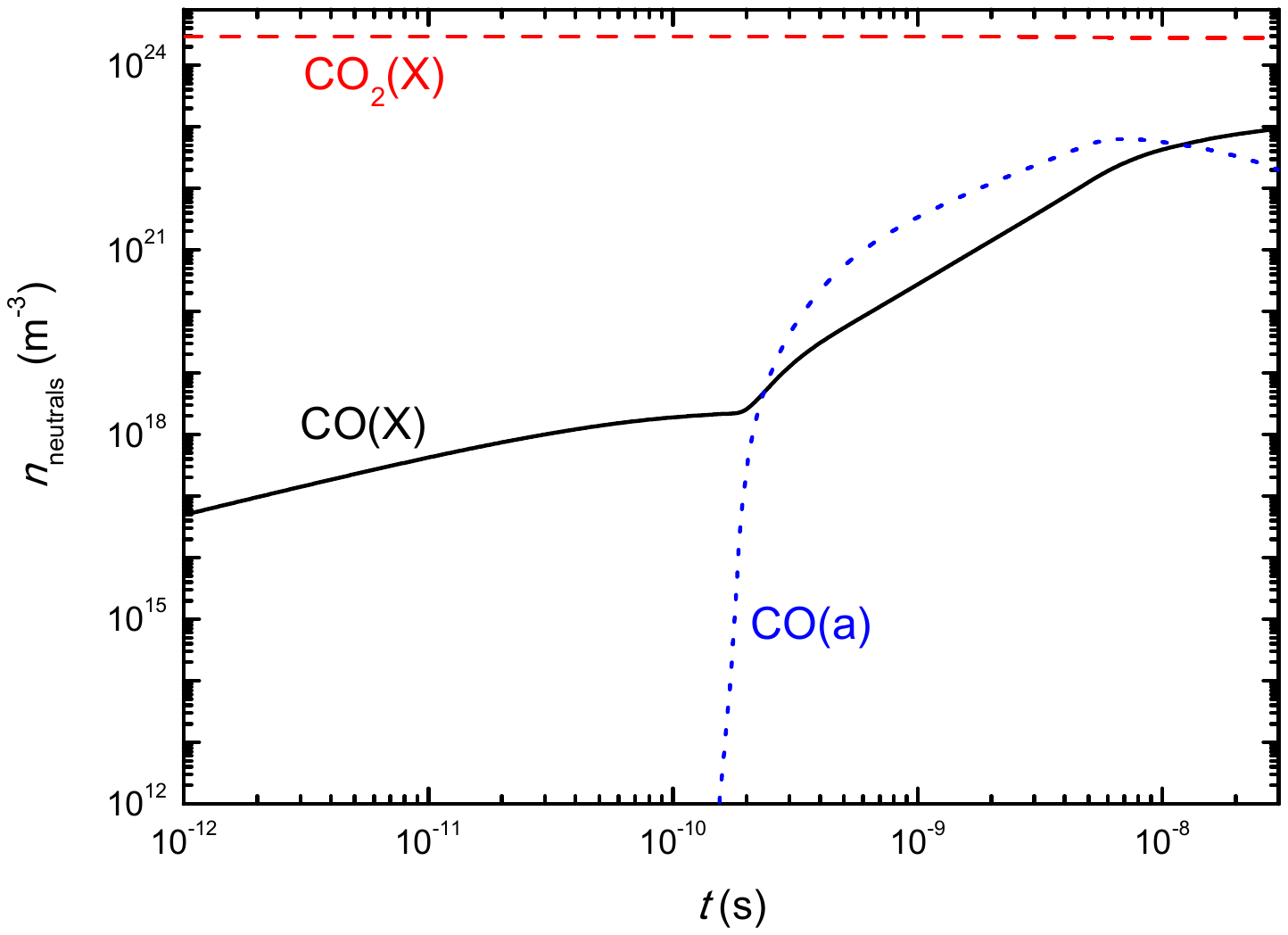}}
	\caption{(a) Temporal evolution of $E/N$ in a pulsed CO$_2$ discharge at $p=1$~atm, $T_g=2500$~K and a gas flow rate of $10$~sccm. The discharge current is prescribed as a triangular pulse with a peak value of $100$~A (red curve). Results are shown for three initial ionization degrees: $n_e/N(0) = 10^{-3}$ (dotted), $10^{-4}$ (dashed) and $10^{-5}$ (solid). (b) Temporal evolution of the concentrations of CO$_2$(X) (dashed curve), CO(X) (solid), and CO(a) (dotted) for the pulsed CO$_2$ discharge shown in panel (a), calculated for an initial ionization degree of $n_e/N(0) = 10^{-4}$.}
	\vspace{-0.3cm}
\end{figure}
In addition, to illustrate the flexibility of LoKI-GM for time-dependent plasma chemistry simulations, figure~\ref{figCO2_03} presents the temporal evolution of the reduced electric field $E/N$ calculated for a pulsed CO$_2$ discharge at $p = 1$~atm, $T_g = 2500$~K and a gas flow rate of $10$~sccm. The discharge current is prescribed as a triangular pulse with a peak value of $100$~A. The simulations were performed for three different initial ionization degrees, $n_e/N(0) = 10^{-3}$, $10^{-4}$ and $10^{-5}$. Since the discharge current is prescribed, the reduced electric field is determined self-consistently from the electron density obtained through the chemistry solution, together with the charge-neutrality condition. As expected, lower initial ionization degrees require higher electric fields to sustain the same current. Consequently, the peak value of $E/N$ increases significantly as $n_e/N(0)$ decreases, exceeding $600$~Td for $n_e/N(0) = 10^{-5}$. As the discharge evolves, however, the electron density rapidly adjusts to the imposed current, causing the reduced electric field to gradually converge toward the same value for all three initial ionization degrees.

To complement the previous results, figure~\ref{figCO2_04} presents the temporal evolution of the concentrations of CO$_2$(X), CO(X) and CO(a), calculated for the case with an initial ionization degree of $n_e/N(0) = 10^{-4}$. The results show that the concentration of CO$_2$(X) remains nearly constant throughout the discharge, corresponding to an overall dissociation fraction of only about $4\%$. In contrast, the concentration of CO(X) increases continuously as a result of electron-impact dissociation and the subsequent chemical conversion of CO$_2$. Starting from values below 10$^{17}$~m$^{-3}$, the CO(X) concentration rises by several orders of magnitude, eventually reaching values on the order of $10^{23}$~m$^{-3}$. 

The electronically excited state CO(a) exhibits a markedly different temporal behavior. Its concentration remains negligible during the initial stages of the discharge and then increases rapidly once the reduced electric field becomes sufficiently high. The production of CO(a) initially proceeds faster than that of ground-state CO(X), leading to a sharp increase in its concentration over several orders of magnitude. This result highlights the potentially important role of electronically excited species not only in low-pressure plasmas~\cite{Liu2025}, where their influence is well established, but also under high-pressure conditions. At later times, however, the growth of the CO(a) population gradually slows down, and its concentration approaches that of CO(X), reflecting the increasing importance of loss mechanisms, including collisional quenching and chemical relaxation.

Overall, these results demonstrate the ability of LoKI-GM to resolve the strongly transient formation of electronically excited species in pulsed discharges while simultaneously capturing their coupling with the evolving plasma chemistry and discharge dynamics.


\subsection{Microkinetic mesoscopic surface model}
\label{showCase:surf}

The recombination of plasma-generated atoms at the walls of a plasma reactor can be described macroscopically through a wall-recombination probability, $\gamma$, which quantifies the fraction of the atomic flux incident on the wall that is lost via recombination and subsequent molecule formation.
The recombination probability $\gamma$ can be estimated from dedicated experimental measurements~\cite{Viegas2025,Viegas2026}. Alternatively, wall recombination can be described through a set of surface reactions with prescribed rate coefficients, an approach that is fully compatible with the LoKI-GM framework.

Coarse-grained descriptions of surface processes, based on the occupation of adsorption sites and on the evolution of surface species densities through reaction kinetics, are commonly referred to as microkinetic mesoscopic models. These models bridge the gap between computationally demanding atomistic simulations and the macroscopic scales relevant to practical plasma systems.

Mesoscopic models can adopt either stochastic~\cite{Guerra2004,Guerra2016,Marinov2017} or deterministic~\cite{Guerra2007,Marinov2017} formulations, whose mathematical treatment can be analytical or numerical. Despite the variety of models and experimental conditions, a broadly accepted picture of atomic wall recombination is that it proceeds through two main mechanisms (see section~\ref{General-rateCoeffVolSurf}):
(i) E-R processes, in which a gas-phase atom (X$_j$) directly recombines with either a chemisorbed atom (X$_s$) or a physisorbed atom (X$_f$). These mechanisms are therefore first-order with respect to the adsorbed atom density;
(ii) L-H processes, in which a physisorbed particle diffuses along the surface and recombines with another adsorbed particle. These mechanisms are consequently second-order with respect to the adsorbed atom density.

An example of LoKI-GM simulations applied to a microkinetic mesoscopic surface model is presented in this section. The test case is based on the work of Guerra~\cite{Guerra2007}, which introduced an analytical model for heterogeneous atomic recombination on silica-like surfaces. The model builds upon the set of reactions originally proposed by Kim and Boudart~\cite{Kim1991}, summarized in table~\ref{tab:surfreac} (reactions (R1) to (R6)) for nitrogen recombination, and proceeds with the analytical derivation of asymptotic expressions for $\gamma$, which can be compared with experimental estimates. The solution of the equation system requires the set of parameters listed in table~\ref{tab:surfparam}.
In the present work, the analytical results reported by Guerra~\cite{Guerra2007} are reproduced numerically using LoKI-GM, employing the rate coefficients and the parameter set of tables~\ref{tab:surfreac}-\ref{tab:surfparam}. In addition, the original model of~\cite{Guerra2007} is extended to include E-R recombination between a gas-phase atom and a physisorbed atom, as well as L-H recombination between two physisorbed atoms.

In table~\ref{tab:surfreac}, the surface densities [F]$_{\rm surf}$ $\equiv$ [F] $(V/A)$ and [S]$_{\rm surf}$ $\equiv$ [S] $(V/A)$ are introduced in order to express the rates of reactions (R1)–(R8) in SI units of m$^{-2}$~s$^{-1}$, thereby rendering the model independent of the specific geometrical factor $V/A$. Moreover, all energy barriers are expressed in SI units of kJ~ mol$^{-1}$ (see table~\ref{tab:surfparam}), and therefore the exponential terms appearing in table~\ref{tab:surfreac} are written using the universal gas constant $R$ instead of the Boltzmann constant $k_B$. With these modifications, the rate coefficients associated with physisorption (R1), desorption (R2), chemisorption (R3) and E-R recombination (R4), (R7) become fully consistent with~\eqref{krj-surface}, by taking $E_i=0$ and $k_i^0=P_i^0$ ($i=1,3,4,7$); adopting $\tau^{-1}_{\rm transp} \simeq \tau^{-1}_{\rm wall} \simeq (v_{th}/4) (A/V)$; and using the near-wall temperature $T_{\rm nw}$ for adsorption processes. 

\begin{table}[h]
\centering
\begin{tabular}{|c|c|c|}
\hline
 Nb. & Reaction & Rate coefficient [m$^{3}$~s$^{-1}$ or m$^{2}$~s$^{-1}$ or s$^{-1}$] \\ 
\hline
R1 & N(g) + F$_v$ $\rightarrow$ N$_f$ & $k_1 = {P_1^0 \over \mathrm{[F]_{surf} + [S]_{surf}}} \, {v_{th}(T_{nw}) \over 4}$ \\
R2 & N$_f$ $\rightarrow$ N(g) + F$_v$ & $k_2 = \nu_d \, \mathrm{exp}\left(-{E_d \over R T_w}\right)$  \\
R3 & N(g) + S$_v$ $\rightarrow$ N$_s$ & $k_3 = {P_3^0 \over \mathrm{[F]_{surf} + [S]_{surf}}} \, {v_{th}(T_{nw}) \over 4}$  \\
R4 & N(g) + N$_s$ $\rightarrow$ N$_2$(g) + S$_v$ & $k_4 = P_{ER}^s \times k_3, P_{ER}^s = P_4^0 \, \mathrm{exp}\left(-{E_{ER}^s \over R T_{nw}}\right)$  \\
R5 & N$_f$ + S$_v$ $\rightarrow$ F$_v$ + N$_{s}$ & $k_5 = {3 \over 4} \times k_D$ (eqs. \ref{eq:kD} and \ref{eq:kD'})   \\
R6 & N$_f$ + N$_s$ $\rightarrow$ N$_2$(g) + F$_{v}$ + S$_{v}$ & $k_6 = P_{LH}^{s} \times k_D, P_{LH}^{s} = P_6^0 \, \mathrm{exp}\left(-{E_{LH}^{s} \over R T_w}\right)$  \\
\hline
R7 & N(g) + N$_f$ $\rightarrow$ N$_2$(g) + F$_{v}$ & $k_7 = P_{ER}^f \times k_3, P_{ER}^f = P_7^0 \, \mathrm{exp}\left(-{E_{ER}^f \over R T_{nw}}\right)$  \\
R8 & N$_f$ + N$_f$ $\rightarrow$ N$_2$(g) + F$_{v}$ + F$_{v}$ & $k_8 = 2 P_{LH}^{f} \times 
{\nu_D \, \mathrm{exp}\left(-{E_D \over R T_w}\right) \over \mathrm{[F]_{surf} + [S]_{surf}}}, P_{LH}^{f} = P_8^0 \, \mathrm{exp}\left(-{E_{LH}^{f} \over R T_w}\right)$  \\
\hline
\end{tabular}
\caption{List of surface reactions describing the recombination of atomic nitrogen on silica-like surfaces, based on the work of~\cite{Guerra2007}.
[F,S]$_{\rm surf}$ $\equiv$ [F,S] $(V/A)$, $R$ is the ideal gas constant and $v_{th}$ is the thermal velocity of gas-phase N (N(g)) to the wall at temperature $T_{nw}$.
}
\label{tab:surfreac}
\end{table}

However, the expression adopted here for the surface diffusion rate coefficient $k_D$, associated with reactions (R5) and (R6) (see table~\ref{tab:surfreac}), differs from that given in equations~\eqref{krj-surface}-\eqref{surfacetime}. In the present formulation, $k_D$ is defined as
\begin{subequations}
\begin{equation}
\label{eq:kD}
k_D = {k_D' \, \nu_d \, \mathrm{exp}\left(-{E_d \over R T_w}\right) \over \mathrm{[S]_{surf}}}
\end{equation}
\begin{equation}
\label{eq:kD'}
k_D' = max\left(0, min\left(1, {\mathrm{[S]} \over \mathrm{[F]}} \left[{\nu_D \over \nu_d} \mathrm{exp} \left({E_d - E_D \over R T_w}\right) - {1 \over 4}\right]\right)\right).
\end{equation}
\end{subequations}
This difference arises from the distinct treatments of surface diffusion adopted in the works of~\cite{Guerra2007} and~\cite{Marinov2017}, the latter of which motivated the formulation presented in equations~\eqref{krj-surface}-\eqref{surfacetime} of section~\ref{General-rateCoeffVolSurf}. These references should be consulted for a complete derivation of the corresponding expressions. Nevertheless, a qualitative comparison of the two approaches is outlined here.

In the model proposed by~\cite{Guerra2007}, collection zones are defined around each chemisorption site. Atoms that become physisorbed within these zones may subsequently migrate to a chemisorption site. The probability that physisorption occurs depends on the relative size of the collection zone: it approaches unity when the collection-zone diameter exceeds the distance between chemisorption sites, and is zero when the collection-zone diameter becomes smaller than that of a physisorption site. In addition, only $1/4$ of the atoms entering the collection zone are assumed to reach the chemisorption site, while the remaining $3/4$ diffuse away in other directions.

Conversely, \cite{Marinov2017} argues that a Monte Carlo formulation provides a more accurate description of surface diffusion and therefore adopts simplified deterministic expressions inspired by Markov-process theory. These formulations were considered sufficiently accurate for surface-kinetics modeling and have consequently been employed in several of our previous works~\cite{Guerra2016,Alves2018,Guerra2019,Afonso2024,Viegas2024}.

\begin{table}[h]
\centering
\begin{tabular}{|c|c|c|}
\hline
 Parameter & Description & Value \\ 
\hline
[N(g)] & Density of atomic nitrogen in gas phase & $10^{21}\ \mathrm{m}^{-3}$ \\
$\mathrm{[F]_{\rm surf}}$ & Density of physisorption sites & $10^{20}\ \mathrm{m}^{-2}$ \\
$\mathrm{[S]_{\rm surf}}$ & Density of chemisorption sites & $2 \times 10^{17}\ \mathrm{m}^{-2}$ \\
$\varphi$ & Fraction of chemisorption sites & $2 \times 10^{-3}$ \\
$P_1^0$ & Sticking probability for physisorption & 1  \\
$P_3^0$ & Sticking probability for chemisorption & 1  \\
$P_4^0, P_{6}^0, P_7^0, P_8^0$ & Pre-exponential factors for recombination & 1  \\
$\nu_d$ & Pre-exponential factor of desorption frequency & $10^{15}\ \mathrm{s}^{-1}$ \\
$\nu_D$ & Pre-exponential factor of diffusion frequency & $10^{13}\ \mathrm{s}^{-1}$ \\
$E_d$ & Desorption barrier & $51$ kJ$~$mol$^{-1}$ \\
$E_D$ & Diffusion barrier & $20.5$ kJ$~$mol$^{-1}$ \\
$E_{ER}^s$ & Recombination barrier for E-R with N+N$_{s}$ & $14$ kJ$~$mol$^{-1}$ \\
$E_{LH}^s$ & Recombination barrier for L-H with N$_{f}$+N$_{s}$ & $14$ kJ$~$mol$^{-1}$ \\
$E_{ER}^f$ & Recombination barrier for E-R with N+N$_{f}$ & 0 \\
$E_{LH}^{f}$ & Recombination barrier for L-H with N$_{f}$+N$_{f}$ & 0 \\
\hline
\end{tabular}
\caption{Surface kinetic parameters, employed in the rate coefficient expressions of table~\ref{tab:surfreac} for the recombination of atomic nitrogen on silica-like surfaces, based on the work of~\cite{Guerra2007}.
}
\label{tab:surfparam}
\end{table}

\begin{figure}[htbp]
\centering
\subfigure[\label{Fig:gammacomp}]{\includegraphics[width=3.0in]{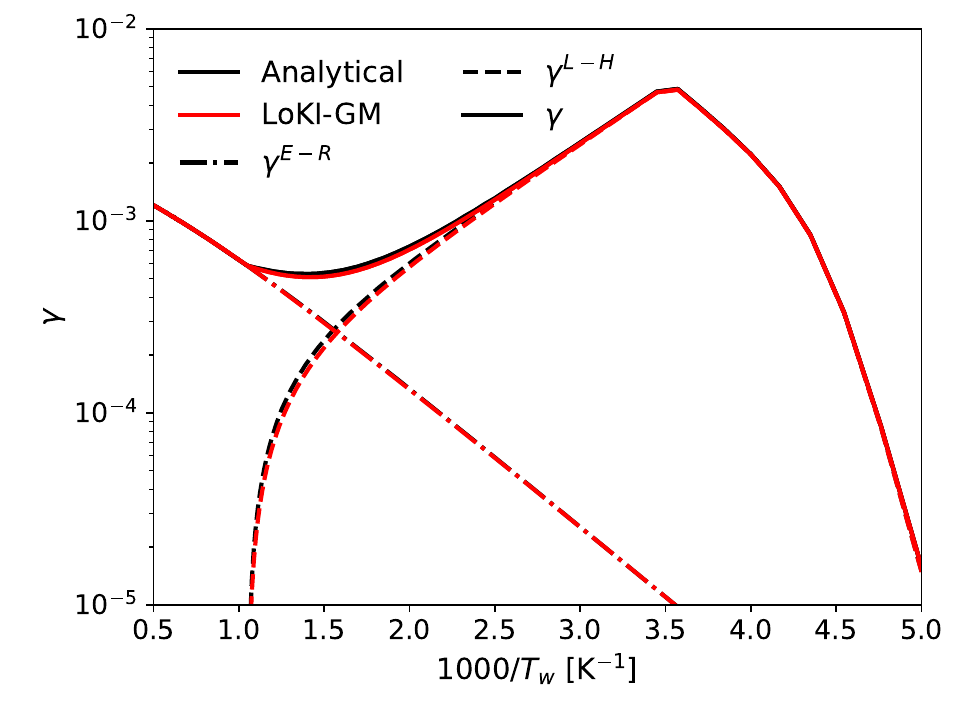}}
\subfigure[\label{Fig:gammaF}]{\includegraphics[width=3.0in]{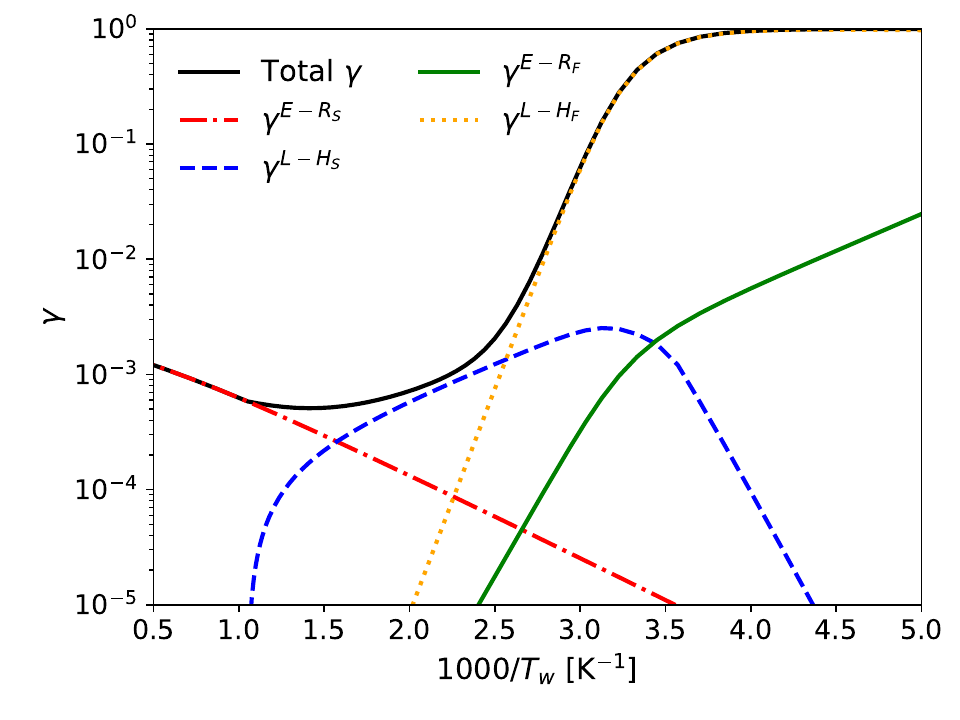}}
\caption{(a) Wall-recombination probability $\gamma$ of N atoms on Pyrex, as a function of the inverse of the near-wall temperature (here, $T_{nw} = T_w$ ranges between $200-2000$~K), from the analytical model of~\cite{Guerra2007} and the corresponding results obtained using LoKI-GM. The individual contributions of reactions (R4) (E-R$_s$) and (R6) (L-H$_s$) are discriminated.
(b) Results obtained with LoKI-GM when including, in addition, reactions (R7) (E-R$_f$) and (R8) (L-H$_f$).
}
\label{Fig:gammas}
\end{figure}

Figure~\ref{Fig:gammacomp} compares the wall-recombination probability $\gamma$ obtained from the analytical model of~\cite{Guerra2007} with the values calculated using LoKI-GM, as function of the wall temperature $T_w$, for the set of reactions (R1)-(R6) and under the assumption that the near-wall gas temperature $T_{nw}$ is equal to $T_w$.

For each case, the individual contributions of the different recombination reactions to $\gamma$ are discriminated. The analytical expression for $\gamma$ is derived in~\cite{Guerra2007} (see equations (73) and (85) therein). Numerically, $\gamma$ is given by the ratio between the total recombination loss rate of atomic nitrogen, $L_\mathrm{rec}^\mathrm{N}$ (in units of m$^{-2}~$s$^{-1}$), and the thermal flux of N, $\phi_\mathrm{N}$, such that
\begin{equation}
\gamma = {L_\mathrm{rec}^\mathrm{N} \over \phi_\mathrm{N}} = 
{L_\mathrm{rec}^\mathrm{N} \over \mathrm{[N(g)]}} {4 \over v_{th}}
\end{equation}

The figure shows an almost exact agreement between the analytical and numerical results.
Only a slight discrepancy is observed in the L-H contribution for wall temperatures above $300$~K, due to approximations adopted in the analytical derivation of $\gamma^{L-H}$.
Furthermore, figure~\ref{Fig:gammacomp} shows that E-R recombination becomes the dominant mechanism at high wall temperatures, where the recombination energy barrier can be more easily overcome. The results also confirm that the corresponding contribution $\gamma^{E-R}$ decreases exponentially with $1000/T_w$.

Low values of $T_w$ also limit desorption, leading to high populations of physisorbed atoms. As a result, the contribution of L-H recombination increases with $1000/T_w$ at low wall temperatures (below $280$~K). The quantity $\gamma^{L-H}$ reaches a maximum around $T_w \simeq 280$~K, due to the increase in the collection-zone diameter as $T_w$ decreases. Once this diameter exceeds the average distance between chemisorption sites, a maximum value of $k_D'$ is obtained (see~\eqref{eq:kD'}). For lower values of $T_w$, $\gamma^{L-H}$ starts to decrease because physisorbed atoms become progressively less able to overcome the desorption, diffusion, and recombination energy barriers through their translational energy.

In table~\ref{tab:surfreac}, the possibility of E-R recombination between a gas-phase atom and a physisorbed atom is included with the addition of (R7). Likewise, L-H recombination between two physisorbed atoms is added through (R8). These reactions are absent from the original mechanisms proposed by~\cite{Guerra2007} and~\cite{Kim1991}. The relevance of (R8) was pointed out in~\cite{Rakhimova2009} and~\cite{Lopaev2011}, and indeed it has been already included in several stochastic simulations~\cite{Guerra2004,Guerra2016}.
Similarly, reaction (R7), corresponding to E-R recombination involving weakly-bound (physisorbed) atoms, was put in evidence by~\cite{Lopaev2011} and~\cite{Booth2019}, and was subsequently incorporated into the wall-recombination mechanism proposed by~\cite{Viegas2024} for oxygen.

In the present work, reactions (R7) and (R8) are included assuming barrier-free recombination of weakly bound atoms. Moreover, for (R8) it is assumed that either of the reacting atoms can diffuse towards the other, which justifies the factor of 2 included in the corresponding rate coefficient (see table~\ref{tab:surfreac}). Since the formulation given by equations~\eqref{eq:kD}-\eqref{eq:kD'}, which rely on the concept of collection zones around chemisorption sites, is not fully applicable to occupied physisorption sites, the rate coefficient for (R8) is instead described using the simpler expressions~\eqref{krj-surface}-\eqref{surfacetime} of section~\ref{General-rateCoeffVolSurf}. 

The inclusion of (R7) and (R8) in this work is intended to illustrate the possible influence of E-R and L-H recombination reactions involving physisorbed atoms. However, this demonstration should be regarded as merely qualitative, since the impact of these reactions is intentionally maximized through several simplifying assumptions: the complete neglect of recombination energy barriers; the absence of removal mechanisms for physisorbed atoms other than desorption and recombination; and the assumption of a constant gas-phase atomic nitrogen density, which would in practice decrease at low $T_w$, due to the high fractional coverage of surface sites and the high recombination probability.

In this framework, figure~\ref{Fig:gammaF} illustrates the potential impact that barrier-free recombination of physisorbed atoms, via reactions (R7) and (R8), may have on the surface recombination probability $\gamma$. At high wall temperatures ($T_w \geq 500$~K, where the coverage of physisorption sites remains low, these reactions have a negligible effect.
However, their contribution becomes increasingly significant as the wall temperature decreases. In particular, L–H recombination between physisorbed atoms becomes the dominant recombination pathway for $T_w \leq 400$~K.
Finally, for $T_w \leq 300$~K, a temperature range that is uncommon in plasma–surface interaction studies, (R7) and (R8) may increase $\gamma$ to values exceeding the typical $10^{-4} - 10^{-2}$ range reported for nitrogen recombination on silica-like surfaces.

\section{Final remarks and outlook} 
\label{Final}

This work introduced the LisbOn KInetics Global Model (LoKI-GM), an open-source code that solves the global kinetic model for pure gases and gas mixtures, providing a self-consistent description of the chemical kinetics and transport of charged and neutral species, in both the volume and surface phases, for user-defined operating conditions: gas composition mixture, pressure, reactor dimensions, and excitation modes (DC/HF glow discharges, post-discharges, and pulsed operation). LoKI-GM handles simulations in any atomic/molecular gas mixture, considering collisions with any volume/surface target states, featuring any type of internal degrees of freedom (electronic, vibrational and rotational), and has been tested under many different plasma conditions.

LoKI-GM comprises two modules, that can run self-consistently coupled or as standalone tools: (i) a Boltzmann solver (LoKI-B)~\cite{Tejero2019,Tejero2021}, which solves the space independent form of the two-term EBE to calculate the isotropic and the anisotropic parts of the electron distribution function, and the corresponding electron macroscopic parameters; (ii) a Chemical solver (LoKI-C), which solves the system of zero-dimensional (volume average) rate balance equations for the main charged and neutral species in the plasma, to calculate the particle densities of the different gas/plasma/surface species, corresponding creation/destruction reaction rates, and the reduced electric field (and any related quantity, such as the discharge current or the discharge power-density). 

LoKI-GM is implemented using a flexible and extensible object-oriented architecture in MATLAB, and supports input/output formats in both graphical user interfaces (presently, the input GUI is limited to LoKI-B simulations) and text-based files in ASCII, JSON and HDF5. The framework uses several modules to describe the mechanisms (collisional, radiative and transport) controlling the creation/destruction of species, includes a gas/plasma thermal model for the self-consistent calculation of the gas temperature, and supports multi-component mean-field microkinetic mesoscopic models to handle surface kinetics in a fully coupled way with volume kinetics. Note that the inclusion of several transport models and support for surface kinetics models are distinguishing features of LoKI-GM compared with other global chemistry models.

Currently, LoKI-GM supports the following simulation types: \textit{steady-state simulations}, in which LoKI-C and LoKI-B are run in a coupled manner; \textit{quasi-stationary simulations}, in which LoKI-C runs as a standalone tool; and \textit{post-discharge simulations}.

In steady-state simulations, used to model active plasmas, LoKI-C updates at steady-state the values of the reduced electric field, the populations of the various heavy species and the gas temperature, which are then supplied to LoKI-B for the solution of the electron kinetics. The numerical workflow involves two (or three) iterative cycles: (i) the \textit{pressure cycle} (optional), over the initial gas density, to obtain the user-prescribed pressure. Note that the steady-state pressure can alternatively be obtained in articulation with the gas inflow/outflow conditions. In this case, the pressure cycle is not needed, as the steady-state pressure is maintained self-consistently during the solution of the rate-balance equations\ref{rate-balance-eq}, by defining a rate coefficient for the outflow under an isobaric condition; (ii) the \textit{neutrality cycle}, over the reduced electric field, to satisfy plasma neutrality, for a user-prescribed electron density, discharge current or discharge power density; and (iii) the \textit{global cycle}, over the densities of the main excited states influencing the EBE, to achieve global convergence of the EEDF and the electron macroscopic parameters.

In quasi-stationary simulations, the electron kinetics is represented by precomputed lookup tables that provide electron swarm parameters, rate coefficients, and power transfer as functions of the electric field (LFA) or the electron mean energy (LEA). The values in these tables are used by the code to calculate the time-dependent populations of the various heavy species, while accounting for the prescribed temporal evolution of the electric field (or the discharge current or the discharge power-density), specified by the user. The
calculations impose the charge-neutrality condition at each instant in time. 

In post-discharge simulations, LoKI-GM first computes the corresponding steady-state solution, followed by a standalone LoKI-C simulation at $E/N = 0$ to describe the time-dependent post-discharge, under the assumption of quasi-neutrality.

This work has also presented several unpublished simulation results obtained with LoKI-GM under a variety of operating conditions.

Simulations in oxygen plasmas analyzed (i) DC glow discharges at gas pressures of $0.2 - 10$~Torr, discharge currents of $10 - 40$~mA and gas flow rates varying linearly from $2$~sccm at $0.2$~Torr to $10$~sccm at $10$~Torr; (ii) HF discharges operating at an excitation frequency of $2.45$~GHz under conditions similar to those of the DC discharges, namely the same discharge power-density. The results confirmed the critical role of the wall-recombination of O($^3$P) in oxygen plasma kinetics; the importance of the enthalpy-exchange in wall-recombination reactions and of the V–T energy transfer in collisions involving O-atoms, for the self-consistent determination of $T_g$; and the increase in the electron density of HF discharges relative to DC discharges operating at the same discharge power-density, resulting from the lower effective reduced maintenance field, $E_{\rm eff}/N$.

Simulations in CO$_2$ plasmas demonstrated (i) the flexibility of LoKI-GM to implement different gas-flow models, ranging from closed reactors to isobaric systems and continuously fed reactors; and (ii) its ability to resolve the strongly transient formation of electronically excited species in pulsed discharges, while simultaneously capturing their coupling with the evolving plasma chemistry and discharge dynamics. The analysis focused on the temporal evolution of the densities of CO$_2$(X), CO(X) and CO(a). In case (i), the simulations considered a DC glow discharge at $p = 5$~Torr, $T_g = 750$~K and $I_ {\rm dc} = 50$~mA, under different gas-flow conditions, highlighting the influence of gas flow on the balance between chemical production and loss processes, as well as on the CO$_2$(X) conversion. Case (ii) addressed pulsed CO$_2$ discharges at $p = 1$~atm, $T_g = 2500$~K and a gas flow rate of $10$~sccm, excited by a triangular discharge-current pulse with a peak value of $100$~A. The simulations showed that lower initial ionization degrees require higher reduced electric fields to sustain the prescribed current. They also showed that the concentration of CO(X) increases continuously as a result of electron-impact dissociation followed by the chemical conversion of CO$_2$, whereas the concentration of CO(a) remains negligible during the initial stages of the discharge and then rises rapidly once the reduced electric field becomes sufficiently high.

We have also presented LoKI-GM simulations applied to a microkinetic mesoscopic surface model, to evaluate the wall-recombination probability $\gamma$ of nitrogen atoms on the walls of a plasma reactor. In this case LoKI-C was used as a standalone solver, and the numerical results are in almost perfect agreement with the analytical model of~\cite{Guerra2007}, over the entire range of wall temperatures investigated. The influence of barrier-free recombination involving physisorbed nitrogen atoms was also analyzed, revealing that L–H recombination between physisorbed atoms becomes the dominant recombination pathway for $T_w \leq 400$~K. The simulations further show that, at lower wall temperatures, the combined contributions of the L-H and E-R mechanisms can increase $\gamma$ above $10^{-2}$.  

These examples were intended to illustrate the main capabilities of LoKI-GM and its flexibility for plasma chemistry studies. The code can also be configured for other applications, \textit{e.g} standalone electron-kinetics simulations (stationary or time-dependent) using LoKI-B~\cite{Tejero2019,Tejero2021}, or plasma chemistry simulations involving only heavy species using LoKI-C as a standalone solver. Additional capabilities and configurations are expected to be incorporated as part of the continued development of LoKI-GM as open-source framework.


\section{Acknowledgments}
The authors acknowledge Dr. T.~C. Dias and MSc. A. Gonçalves for their support and dedication to the development of LoKI-GM.  \\
This work was supported by the FCT - Fundação para a Ciência e Tecnologia, I.P., under projects \\
UID/50010/2025 (https://doi.org/10.54499/UID/50010/2025), \\
UID/PRR/50010/2025 (https://doi.org/10.54499/UID/PRR/50010/2025), \\
UID/PRR2/50010/2025 (https://doi.org/10.54499/UID/PRR2/50010/2025), \\
LA/P/0061/2020 (https://doi.org/10.54499/LA/P/0061/2020), \\
UID/04650/2025 (https://doi.org/10.54499/UID/04650/2025) and \\
2022.04128.PTDC (doi.org/10.54499/2022.04128.PTDC).

\end{document}